# Gas temperature measurement based on contrast reversal in mid-infrared $CO_2$ images

**HIDEKI T. MIYAZAKI,* TAKESHI KASAYA, MASAHIRO SAITO, KAZUYA KIMOTO, YUTARO TSUIKI, AND TETSUYUKI OCHIAI**

*National Institute for Materials Science (NIMS), Tsukuba, Ibaraki 305-0047, Japan*
*MIYAZAKI.Hideki@nims.go.jp

**Abstract:** We demonstrate noninvasive measurement of gas temperature based on the optical gas imaging. Gas flows containing carbon dioxide ($CO_2$) appear as either bright or dark images, depending on the relative temperatures of the background and the gas, when using a narrowband mid-infrared camera tuned to the $CO_2$ absorption wavelength at 4.3 μm. When the background temperature is varied continuously, the gas image vanishes transiently and then the contrast reverses. The specific background temperature at the point when the gas image disappears provides the gas temperature. This technique is an evolved implementation of the classical line reversal method, made possible by advanced infrared devices. We also apply this technique to two-dimensional temperature mapping and to dynamic emissions from engine exhaust and human breathing.

## 1. Introduction

The measurement of gas temperature continues to be a challenging problem [1,2] due to the poor thermal conductivity and boundary heat transfer of gases. The most straightforward method is to expose a solid probe, such as a thermocouple, to the gas. However, in using this simple method, it is difficult to obtain reliable results due to thermal conduction through the probe and radiation from the probe's surface. Therefore, various noninvasive methods have been developed, such as infrared radiation computed tomography [3], Rayleigh [4] or Raman scattering [5], and coherent anti-Stokes Raman spectroscopy [6]. Recently, it has become possible to determine temperature from a single absorption lineshape using quantum cascade lasers [7]. However, these techniques require precise optical systems, making them unsuitable for applications outside the laboratory.

Among the many noninvasive techniques for gas temperature measurement, there is an exceptionally simple method: line reversal [1,2,8]. The line reversal method was widely studied in the early 20th century as a way to determine flame temperature (1000–2800 K). In this approach, a thermal radiation light source with variable temperature (intensity) is placed behind the flame, and its light passing through the flame is observed with a spectrometer while changing the temperature of the light source. The flame temperature is determined as the temperature of the light source when the bright lines turn to dark lines [9,10]. Although the reversal of the D lines of Na atoms is well known [11], the use of absorption lines of carbon dioxide ($CO_2$) and water vapor ($H_2O$) in the mid-infrared region has also attracted interest from an early stage [10,12,13]. In 1928, the temperature of $CO_2$ contained in a flame was measured by the line reversal method in the 4-μm region [14].

Recent advances in infrared detectors have permitted optical gas imaging (OGI) [15,16]. By restricting the observation wavelength to a narrowband tuned to the absorption of a specific gas, that gas can be selectively visualized remotely. Dispersive OGI based on Fourier transform infrared spectroscopy is versatile but slow in image acquisition [17]. In contrast, nondispersive OGI without a spectrometer allows dynamic tracking of specific gas movements [18–21], which is the topic of this study. While the sensitivity band of the imaging device itself is used for visualizing gases in some cases [22], band-pass filters or gas cells matched to the gas of interest are usually employed [20]. In particular, filters cooled at cryogenic temperatures [21] provide images with higher quality than uncooled, room-temperature filters [18,19]. In either case, the use of a cooled camera is necessary for satisfactory results because faint intensity changes in narrow bands must be captured.

OGI technology has been used for detecting hazardous gas leaks [18–20] and visualizing gas flows [21–24]. In addition, since the COVID-19 pandemic, there has been growing interest in visualizing human exhalation [25–29].

This paper presents a noninvasive method for gas temperature measurement based on the principle of line reversal and use of the OGI technique. A temperature-variable thermal radiation light source is placed behind a gas containing $CO_2$, and its light passing through the gas is observed with a $CO_2$ imaging camera while changing the temperature of the light source. In this method, instead of the reversal of line spectra, black and white gas appearances are reversed in two-dimensional (2D) images; here, we call this phenomenon contrast reversal. Accordingly, the gas temperature is determined as the temperature of the light source at the point when the contrast reverses.

A few works have reported on gas temperature measurement using OGI [17,30,31]. However, they all determined temperature based on the assumed theoretical temperature dependence of the gas absorption coefficient. On the other hand, the contrast reversal method is a null method, where the gas temperature is determined as the background temperature at the moment the gas becomes invisible. The accuracy of the measurement is mainly determined by the accuracy of the light source temperature, rather than by the assumed absorption properties of the gas.

The evolution from classical line spectra reversal to 2D image contrast reversal brings strong advantages. With the aid of advanced image processing techniques, the rich information in the images can be fully exploited, permitting accurate temperature determination even from images with poor contrast. Moreover, this approach can obtain 2D temperature distribution. Using a high frame rate, typically 30 frames per second, makes it possible to measure the temperature of time-varying targets such as engine exhaust and human breath. In addition, both the camera and light source are portable, thus facilitating application outside the laboratory.

With the growing demand for reducing $CO_2$ emissions and the transition to new fuels, monitoring gas emissions into the atmosphere is becoming increasingly important. Techniques of quantitatively measuring gas emissions using OGI are thus promising [16,32]. However, because the appearance of gas in OGI strongly depends on the gas temperature [32–34], the first step in quantifying gas emissions is to measure the temperature of a gas, and this challenge is the primary motivation of this work. Because emitted gas contains $CO_2$ in most cases, the method presented here can usually be applied as is; moreover, by seeding emissions with $CO_2$, this method can be applied to any gas.

The remainder of this paper is structured as follows. Section 2 discusses the theoretical foundation of contrast reversal. Section 3 introduces the equipment used as well as the image measurement method. Section 4 verifies the basic principle by measuring gases with known temperatures. Section 5 applies this technique to target gases of unknown temperature. Section 6 provides a discussion and summary. Further details are discussed in Supplementary Documents.

## 2. Fundamentals of contrast reversal in gas imaging

### 2.1 Radiative transfer equation and blackbody radiation

In this section, we clarify how the temperature of a gas is reflected in infrared optical images. Suppose that radiation from a light source at temperature $T_s$ passes through a gas at temperature $T_g$ and is detected by a camera (Fig. 1(a)). The spectral radiation intensity $i_\nu$ (W/(m$^2$ sr cm$^{-1}$)) at frequency (wavenumber) ν (cm$^{-1}$) along the coordinate $z$ from the light source to the camera follows the radiative transfer equation:

$$\frac{\mathrm{d}i_\nu}{\mathrm{d}\kappa_\nu} = -i_\nu(\kappa_\nu) + i_\nu^{\mathrm{source}}(\kappa_\nu). \tag{1}$$

Here, $\kappa_\nu$ is the optical thickness (dimensionless) and $i_\nu^{\mathrm{source}}(\kappa_\nu)$ is the source function representing the radiation from the gas. We basically follow the definitions of Siegel and Howell [35] and the HITRAN database [36]. Gas scattering can be neglected [23]. Furthermore, stimulated emission can be ignored in the range $T_\mathrm{g}$ = 0–200°C, which is used in this work. Throughout this study, pressure is assumed to be constant (standard pressure: 101,325 Pa).

The optical thickness is

$$\kappa_\nu = \int_0^\infty a_\nu \,\mathrm{d}z = \int_0^\infty k_\nu n c \mathrm{d}z, \tag{2}$$

where $a_\nu$ (cm$^{-1}$) is the absorption coefficient, $k_\nu$ (cm$^2$/molecule) is the absorption cross section, $n = T^0 N_\mathrm{A}/(TV^0)$ (molecules/cm$^3$) is the total volume number density of the gas, $T^0$ is the standard temperature (273.15 K), $V^0$ is the volume occupied by 1 mol of gas at standard state (2.24×10$^4$ cm$^3$), $N_\mathrm{A}$ is the Avogadro constant, $T$ is temperature, and $c$ is the mole fraction (concentration) of the gas. As an indicator of gas concentration, the column number density $u$ (molecules/cm$^2$) is considered [36]:

$$u = \int_0^\infty n c \mathrm{d}z.$$

For practical convenience, the product of concentration and thickness ζ (ppm m), referred to simply as the column density, is often used [16]:

$$\zeta = \int_0^\infty c \,\mathrm{d}z.$$

The incident light is assumed to be blackbody radiation at temperature $T_\mathrm{s}$. Since there is no scattering, the source function corresponds to the blackbody radiation of the gas at temperature $T_\mathrm{g}$. The blackbody radiation intensity $i_{\nu\mathrm{bb}}(T)$ (W/(m$^2$ sr cm$^{-1}$)) is given as [37]

$$i_{\nu\mathrm{bb}}(T) = \frac{2hc_0^2\nu^3}{\exp(hc_0\nu/k_\mathrm{B}T)-1},$$

where $h$ is the Planck constant, $c_0$ is the speed of light in vacuum, and $k_\mathrm{B}$ is the Boltzmann constant. The radiative transfer equation (Eq. (1)) can be written in an integrated form using blackbody radiation intensity:

$$i_\nu(\kappa_\nu) = i_{\nu\mathrm{bb}}(T_\mathrm{s}) \exp(-\kappa_\nu) + \int_0^{\kappa_\nu} i_{\nu\mathrm{bb}}\big(T_\mathrm{g}\big)\exp[-(\kappa_\nu - \kappa_\nu^*)]\mathrm{d}\kappa_\nu^*. \tag{3}$$

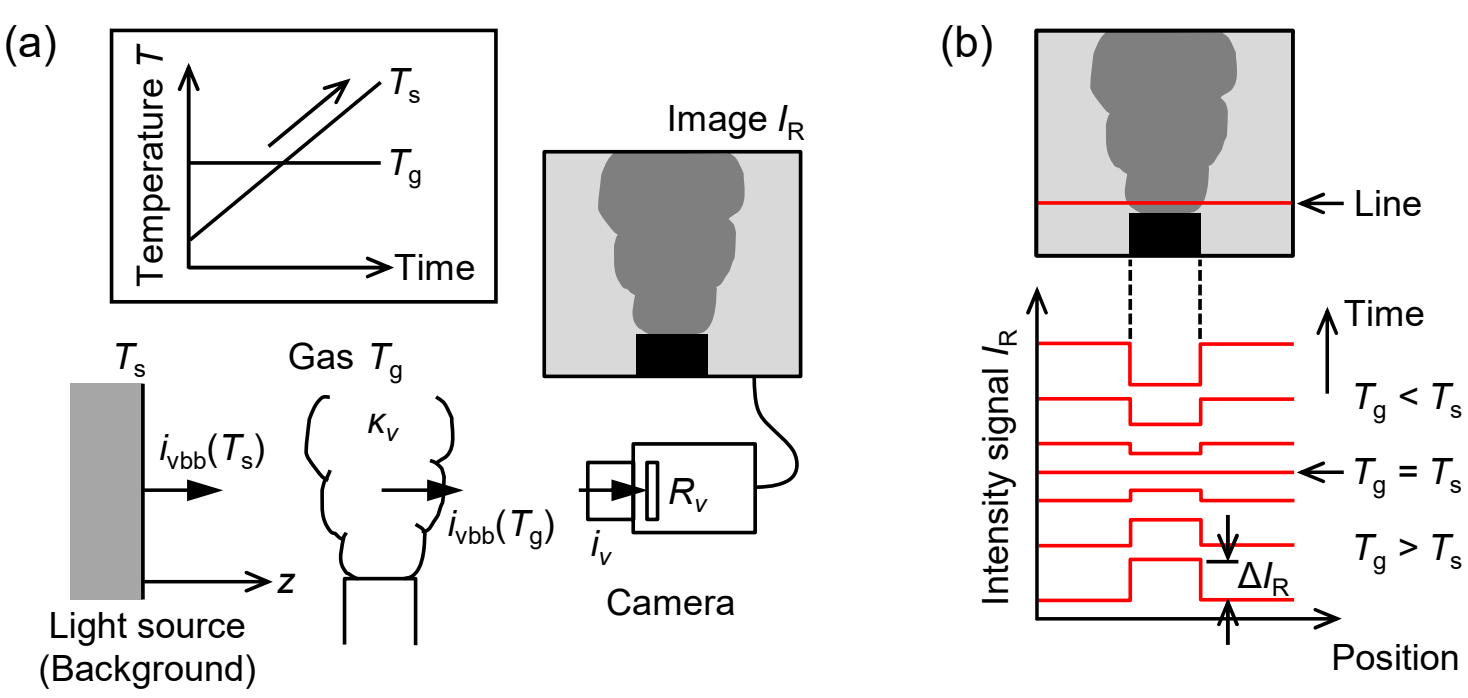

Fig. 1. Overview of gas temperature measurement based on contrast reversal. (a) Proposed system: arrangement of light source, gas, and camera, coordinate system, temperature of each part, and observed image. Inset: variation of light source temperature $T_\mathrm{s}$ crossing the gas temperature $T_\mathrm{g}$. (b) Change in intensity $I_\mathrm{R}$ profile along a line just after the emission in the gas image, according to the change (increase in this figure) of light source temperature $T_\mathrm{s}$.

### *2.2 Radiative transfer at a single frequency in uniform gas*

Let us consider radiative transfer at a single frequency ν for a simple situation, where the temperature and concentration of the gas are uniform and the thickness $L$ (m) is clearly defined. It is assumed that only the target gas exhibits absorption. In this case, Eq. (2) is simplified to $\kappa_\nu = k_\nu ncL = uL$, where $u = ncL$. Equation (3) is also simplified as

$$i_\nu(\kappa_\nu) = \left(1 - A_{\nu g}\right) i_{\nu bb}(T_s) + A_{\nu g} i_{\nu bb}\left(T_g\right), \quad (4)$$

where $A_{\nu g} = 1 - \tau_{\nu g}$ is absorptivity and $\tau_{\nu g} = \exp(-\kappa_\nu)$ is transmissivity.

Typical spectra are shown in Fig. 2. The $CO_2$ absorption cross section $k_\nu$ is a collection of sharp absorption lines as seen in Fig. 2(a). The absorption spectrum for $CO_2$ in this study follows HITRAN [36]. For details on the spectra in Fig. 2 and equations in Sec. 2.2 and 2.3, refer to Supplement 1, Sec. A.

The responsivity $R_\nu$ of the $CO_2$ imaging camera used in this study (Fig. 2(b)) covers a much broader bandwidth, with a full width at half maximum of $\Delta\nu = \nu_2 - \nu_1$, than individual absorption lines. Therefore, the experimentally observed signal is an integration over many absorption lines, as discussed below (Sec. 2.3, Fig. 3(d)). For $T_g$ = 50°C, $T_s$ = 75°C, and several column densities ζ, $i_\nu$ obtained from Eq. (4) is shown in Fig. 2(c); $i_\nu$ for $T_s$ = 25°C is shown in Fig. 2(d).

While column density ζ is a convenient parameter simply based on gas concentration $c$ and thickness, absorptivity cannot be specified with only ζ. The essential parameter defining absorptivity is the column number density $u$. Since $u$ also depends on temperature $T$ due to the $n$ term, absorptivity is determined by both ζ and $T$. Therefore, whenever ζ is specified in this paper, the temperature, at which the value of ζ is defined, is also noted.

When the background is hotter than the gas (Fig. 2(c)), background radiation is absorbed by the gas, and the signal detected by the camera decreases as the gas becomes thicker (higher column density); the gas appears dark in contrast to the background (negative image). However, even when the gas is so dense as to be completely opaque, the detected intensity does not reach zero. The minimum intensity is limited by the blackbody radiation at the gas temperature.

Conversely, when the background is colder than the gas (Fig. 2(d)), the radiation from the gas is added to the background radiation, and the gas appears brighter as it becomes denser (positive image). Here, the maximum brightness is limited by the blackbody radiation at the gas temperature. Accordingly, Fig. 2(c) and (d) show gas images with reversed contrast.

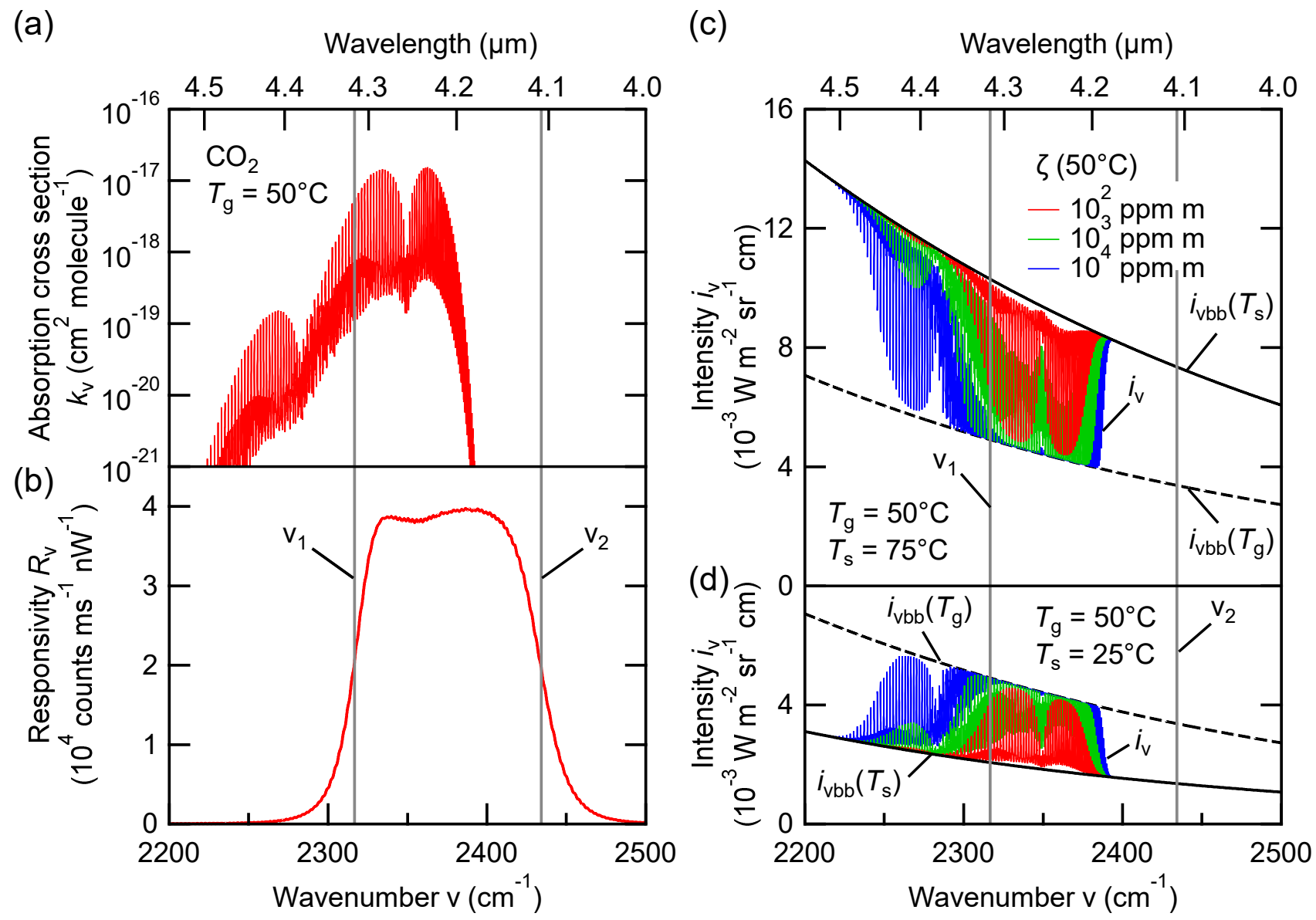

Fig. 2. Spectra related to radiative transfer for gases with uniform temperature and concentration. (a) Absorption cross section spectrum $k_\nu$ of $CO_2$ gas at temperature $T_g$ = 50°C. (b) Responsivity spectrum $R_\nu$ of $CO_2$ imaging camera used in study. (c) Radiation intensity spectrum $i_\nu$ and blackbody radiation intensities $i_{\nu bb}$ at gas temperature $T_g$ = 50°C, light source temperature $T_s$ = 75°C, and column density $\zeta$ (50°C) = $10^2$, $10^3$, $10^4$ ppm m. (d) Similar results at $T_g$ = 50°C and $T_s$ = 25°C. In all panels, vertical lines indicate full width at half maximum of responsivity spectrum $R_\nu$.

### 2.3 Radiative transfer in uniform gas over a finite bandwidth

For a gas with column number density $u$, the intensity signal $I_R(u)$, which is actually detected by the camera, is given by integrating $i_\nu(\kappa_\nu)$ multiplied by the responsivity spectrum $R_\nu$ of the camera (Fig. 2(b)) over $\nu$:

$$I_R(u) = \int_0^\infty R_\nu i_\nu(\kappa_\nu) d\nu$$

$$= \int_0^\infty R_\nu i_{\nu bb}(T_s) \exp(-\kappa_\nu)\, d\nu + \int_0^\infty R_\nu \int_0^{\kappa_\nu} i_{\nu bb}(T_g) \exp[-(\kappa_\nu - \kappa_\nu^*)] d\kappa_\nu^*\, d\nu. \quad (5)$$

For the convenience of practical applications, Eq. (5) can be described in a simpler form, as Eq. (4) for a single frequency case:

$$I_R(u) \approx (1 - A_g) I_{Rbb}(T_s) + A_g I_{Rbb}(T_g). \quad (6)$$

Refer to Supplement 1, Sec. A.3 for the derivation. Here,

$$I_{Rbb}(T) = \int_0^\infty R_\nu i_{\nu bb}(T) d\nu$$

is the signal measured by the detector with responsivity $R_\nu$ for blackbody radiation at temperature $T$, which can be determined by calibration experiments.

$$A_g = \frac{\int_0^\infty R_\nu [1 - \exp(-\kappa_\nu)] d\nu}{\int_0^\infty R_\nu d\nu}$$

is the $CO_2$ absorptivity measured by the detector with responsivity $R_\nu$; this depends not only on the column number density $u$ but also on the gas temperature $T_g$. While $A_g$ can also be obtained experimentally, this study uses the $A_g$ values calculated by HITRAN. $A_g$ based on HITRAN was confirmed to be consistent with experimental results (Fig. S1(a)). Equation (6) is the fundamental formula, and it is used below for practical corrections.

In the absence of gas,

$$I_{\mathrm{R}}(0) = I_{\mathrm{Rbb}}(T_{\mathrm{s}}),$$

and thus the radiation signal difference due to the presence of the gas is

$$\Delta I_{\mathrm{R}} = I_{\mathrm{R}}(u) - I_{\mathrm{R}}(0) \approx A_{\mathrm{g}}\left[I_{\mathrm{Rbb}}(T_{\mathrm{g}}) - I_{\mathrm{Rbb}}(T_{\mathrm{s}})\right]. \tag{7}$$

When the background temperature $T_{\mathrm{s}}$ and the gas temperature $T_{\mathrm{g}}$ are equal, $\Delta I_{\mathrm{R}} = 0$ (i.e., the gas becomes invisible), regardless of $A_{\mathrm{g}}$ (i.e., regardless of the concentration or absorption properties of the gas).

This is demonstrated for $T_{\mathrm{g}}$ = 50°C (as in Fig. 2) and the representative column densities in Fig. 3(a). When the background (light source) temperature $T_{\mathrm{s}}$ is varied, $\Delta I_{\mathrm{R}}(T_{\mathrm{s}}) = 0$ at $T_{\mathrm{s}} = T_{\mathrm{g}}$, regardless of the column density $\zeta$. In this manner, the gas temperature can be determined from images. The $T_{\mathrm{s}}$ that gives $\Delta I_{\mathrm{R}}(T_{\mathrm{s}}) = 0$ is called the contrast reversal temperature $T_{\mathrm{s}}^{\mathrm{rev}}$.

Here, let us summarize the concept of the contrast reversal method with reference to Fig. 1. A planar infrared light source with controllable temperature is placed behind the gas, and the background (light source) is observed through the gas (Fig. 1(a)). We then consider the intensity profile in the image just after the gas emission (Fig. 1(b)). When the light source temperature $T_{\mathrm{s}}$ is varied (e.g., increased) across the expected gas temperature $T_{\mathrm{g}}$ (Fig. 1(a), inset), the image intensity signal $I_{\mathrm{R}}$ basically increases in accordance with $T_{\mathrm{s}}$, but the signal difference of the gas against the background, $\Delta I_{\mathrm{R}}$, changes gradually from positive to negative. During this process, there is a moment when the gas and background intensity match, resulting in a flat profile where the gas becomes invisible. Consequently, the unknown gas temperature $T_{\mathrm{g}}$ can be determined as the light source temperature $T_{\mathrm{s}}^{\mathrm{rev}}$ at which this contrast reversal occurs.

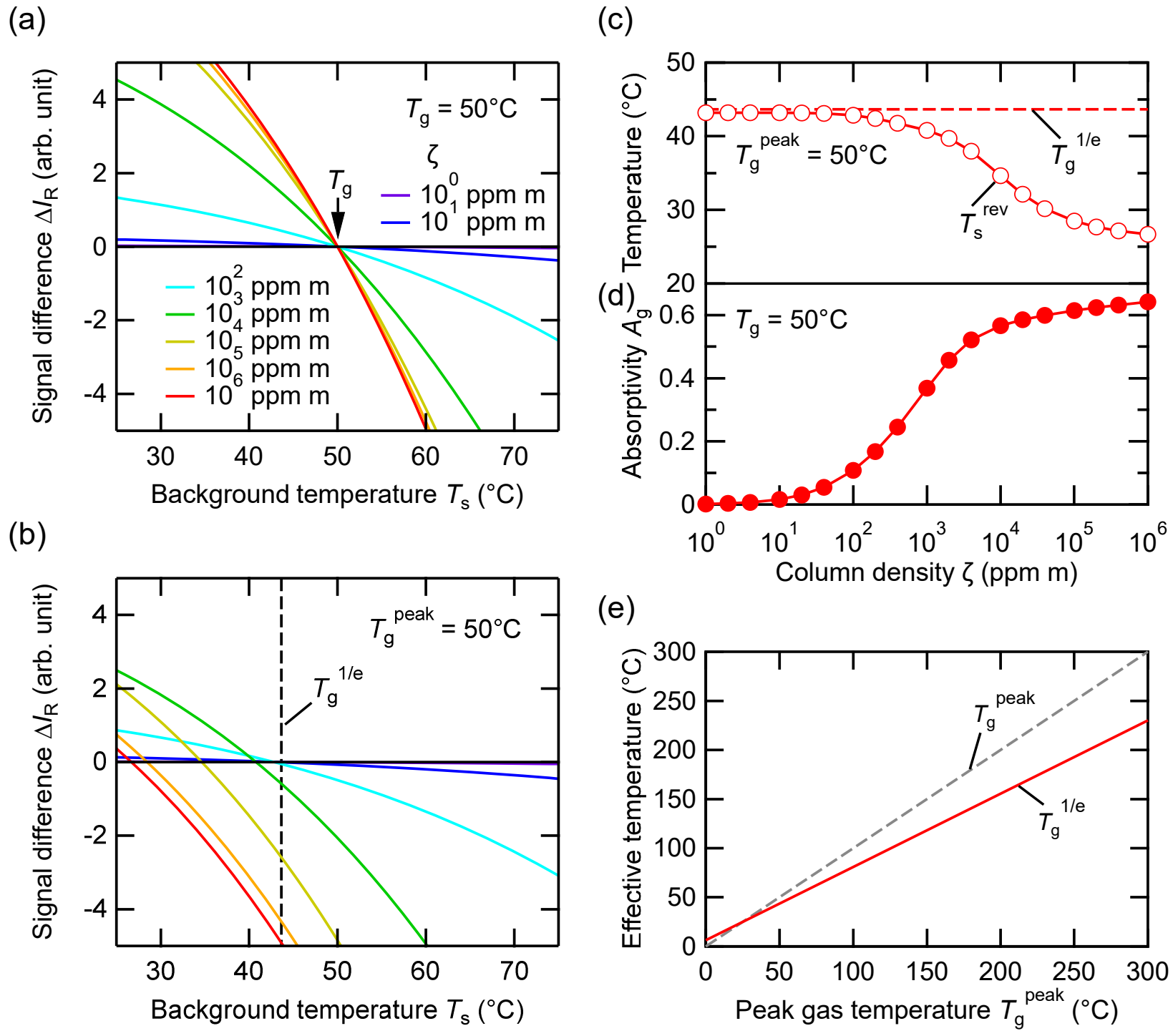

Fig. 3. Contrast reversal behavior for $CO_2$-containing gas at temperature $T_g$ = 50°C observed by a $CO_2$ imaging camera with responsivity $R_\nu$ and a finite bandwidth. (a) Relationship of signal difference $\Delta I_R$ to light source temperature $T_s$ for uniform gas with temperature $T_g$ = 50°C and various column densities $\zeta$ (50°C) shown in legend. (b) $\Delta I_R$–$T_s$ relation for nonuniform gas with Gaussian-distributed temperature and concentration: peak gas temperature $T_g^{peak}$ = 50°C, concentration corresponding to the values of $\zeta$ (25°C) in the legend of (a). (c) Relationship between contrast reversal temperature $T_s^{rev}$ and $\zeta$ for Gaussian-distributed $CO_2$-containing gas with $T_g^{peak}$ = 50°C and concentration corresponding to $\zeta$ in the horizontal axis. The $1/e$-width average temperature $T_g^{1/e}$ is also indicated. (d) Relationship between absorptivity $A_g$ and $\zeta$ (50°C) for $CO_2$-containing uniform gas at $T_g = 50$°C. (e) Relationship between $T_g^{peak}$ and $T_g^{1/e}$ for gas with Gaussian-distributed temperature at ambient temperature $T_a$ = 25°C.

### *2.4 Radiative transfer in nonuniform gas over a finite bandwidth*

The aim of this study is to quantify the temperature of gas plumes released into free space, which would be beneficial for real-world applications. Accordingly, the above discussion must be extended to general cases where the temperature and concentration are not uniform.

Conventional line reversal methods have also been applied to nonuniform gases. Flames in free space are essentially nonuniform; a hot core is surrounded by cold boundary layers. Therefore, the line reversal temperature should be regarded as an average or effective temperature [1,8]. Strictly speaking, the line reversal method is a null technique for determining the effective temperature of flames or gases [38]. Although the relationship between the temperature distribution of a flame and the line reversal temperature has been discussed [39], different temperature distributions were assumed for each case [40,41], and no generalized theory has been found. Here, we consider a gas whose temperature and concentration are distributed in a Gaussian manner along the thickness direction, which would be a sufficiently general assumption.

We assume that the temperature and concentration profiles of the gas are described by respective Gaussian functions with a common width. The gas concentration profile is assumed to have a peak value $c^{peak}$ and to be zero in the environment. The peak concentration is set to achieve a specific value of column density ζ (Fig. S1(c)), assuming a uniform temperature of 25°C. The temperature profile is assumed to have a peak gas temperature $T_g^{peak}$ = 50°C in an ambient temperature $T_a$ = 25°C (Fig. S1(d)). As shown in Eq. (2), the optical thickness $\kappa_\nu$ is determined by $k_\nu$, $n$, and $c$. Since both $k_\nu$ and $n$ depend on $T_g$, $\kappa_\nu$ is influenced by the spatial distribution of $T_g$ as well as that of $c$. Taking into account all of these spatial distributions in Eq. (5), the detected signal $I_R(u)$ for nonuniform gas is obtained for various light source temperatures $T_s$.

The relationship between the signal difference $\Delta I_R$ due to the presence of the gas and the light source temperature $T_s$ is shown in Fig. 3(b) for various column densities ζ. Unlike the uniform case (Fig. 3(a)), the gas becomes invisible at $T_s$ lower than 50°C. Furthermore, the contrast reversal temperature $T_s^{rev}$ decreases as ζ increases. The relationship between $T_s^{rev}$ and ζ is summarized in Fig. 3(c). While ζ is low, $T_s^{rev}$ remains nearly constant and close to the $1/e$-width average temperature $T_g^{1/e}$ (Fig. S1(d)) shown by the dashed line. However, $T_s^{rev}$ starts to decrease at ζ ~ 1000 ppm m, corresponding to the region where $A_g$ begins to saturate (Fig. 3(d)). For a sufficiently dense gas, radiation from the light source is attenuated due to the high opacity of the gas, and only radiation from the surface layer at the exit side of the gas reaches the camera. Consequently, in the high ζ limit, the determined $T_s^{rev}$ approaches the environmental temperature at the edge of the Gaussian distribution.

This behavior is essentially identical to the single-frequency, nonuniform case presented in Fig. S2 (see Supplement 1, Sec. B). From these findings, it seems to be a universal nature that the contrast reversal method gives a temperature value close to $T_{\mathrm{g}}^{1/\mathrm{e}}$ for relatively low-density gases.

In summary, Eqs. (6) and (7) can also be applied to nonuniform gas flows in free space. A nonuniform gas with Gaussian temperature and concentration distributions should be regarded as a uniform gas with an equivalent temperature of $T_{\mathrm{g}}^{1/\mathrm{e}}$. For such nonuniform gases, the contrast reversal temperature $T_{\mathrm{s}}^{\mathrm{rev}}$ gives the $1/e$-width average temperature $T_{\mathrm{g}}^{1/\mathrm{e}}$. The relationship between the measured $T_{\mathrm{g}}^{1/\mathrm{e}}$ and the peak temperature $T_{\mathrm{g}}^{\mathrm{peak}}$ is analytically denoted as $T_{\mathrm{g}}^{1/\mathrm{e}} = 0.746T_{\mathrm{g}}^{\mathrm{peak}} + 0.254T_{\mathrm{a}}$, as shown in Fig. 3(e). For more complicated gas distributions, special considerations are necessary. However, such cases are outside the scope of this study.

### *2.5 Necessary corrections considering the influence of atmospheric $CO_2$*

The $CO_2$ discussed in this study also exists in the atmosphere at approximately 400 ppm, and its absorption cannot be ignored. The properties of the optical system have also been neglected in previous sections. Here, we discuss necessary corrections for gas temperature measurements in real situations. For details, see Supplement 1, Sec. C and Fig. S3.

Two factors must be considered. The first is the radiative transfer in the background space. In all of the above discussions, $I_{\mathrm{Rbb}}(T_{\mathrm{s}})$ should be replaced with $I_{\mathrm{Rbb}}(T_{\mathrm{s}}^{\mathrm{eff}})$, where

$$I_{\mathrm{Rbb}}\left(T_{\mathrm{s}}^{\mathrm{eff}}\right) = \tau_{\mathrm{b}}^{\mathrm{abs}} I_{\mathrm{Rbb}}(T_{\mathrm{s}}) + \left(1 - \tau_{\mathrm{b}}^{\mathrm{abs}}\right) I_{\mathrm{Rbb}}(T_{\mathrm{a}}). \tag{8}$$

Due to the transmissivity $\tau_{\mathrm{b}}^{\mathrm{abs}}$ of the background space between the light source and the gas, the intensity of light that actually irradiates the gas is reduced from the original intensity. Meanwhile, radiation from atmospheric $CO_2$ at temperature $T_{\mathrm{a}}$ is added. As a result, the light incident on the gas corresponds to blackbody radiation at an effective temperature $T_{\mathrm{s}}^{\mathrm{eff}}$.

The second factor is the bandwidth over which transmissivity is defined. Here, the transmissivity $\tau_{\mathrm{b}}^{\mathrm{abs}}$ is for the local band where the camera's sensitivity overlaps the gas absorption band, and the superscript indicates that this is for the absorption band. When considering signal difference $\Delta I_{\mathrm{R}}$ in Eq. (7), which is used to extract the effect of the presence of the gas, the signals in $I_{\mathrm{R}}(u)$ and $I_{\mathrm{R}}(0)$ within the transparent band cancel each other out. Therefore, only the transmissivity within the absorption band must be considered (for detailed derivation, see Supplement 1, Sec. C and Fig. S1(b)). In all subsequent cases, a $\tau_{\mathrm{b}}^{\mathrm{abs}}$ correction determined by the ambient $CO_2$ concentration and background distance was applied to each measurement.

## 3. Experimental methods

### *3.1 $CO_2$ imaging narrowband infrared camera*

For use in our experiments, we adopted a custom-made InSb camera equipped with a built-in cooled filter for $CO_2$, FLIR A6796 (Fig. 4(a) and (b)). Both the image sensor and the filter were cooled to 80 K. The camera's main features included image resolution of 640×512 pixels, frame rate up to 480 fps, and intensity resolution at 14 bits. The responsivity spectrum (Fig. 2(b)) had a center wavelength of 4.21 μm and a full width at half maximum of 0.21 μm. This band was shifted to the shorter wavelength side relative to the $CO_2$ absorption (Fig. 2(a)). Consequently, even for highly dense $CO_2$, a certain amount of incident light could always be detected, ensuring some amount of image output. More detailed information on this camera is provided in Supplement 1, Sec. D. While many OGI cameras with cooled filters are available for hydrocarbon molecules [15], those for $CO_2$ are limited [21,25,26,28,29]. However, with the growing interest in carbon emissions and $CO_2$ storage, the demand is expected to increase.

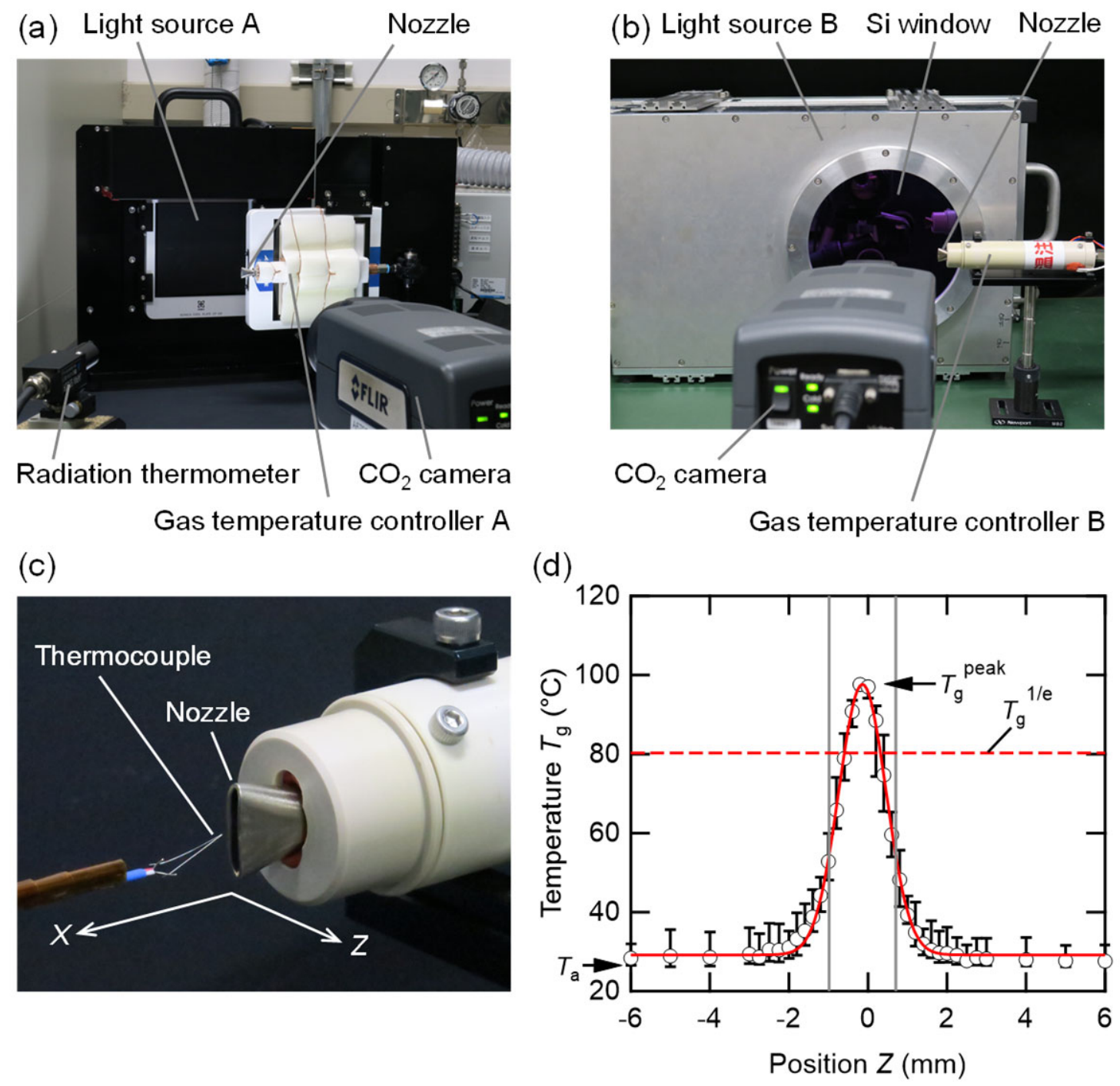

Fig. 4. Measurement system. (a) Setup for measuring gas temperature $T_{\mathrm{g}}$ near room temperature using light source A and gas temperature controller A; both are Peltier-controlled. (b) Setup for measuring a higher gas temperature $T_{\mathrm{g}}$ using vacuum-sealed light source B and gas temperature controller B. (c) Nozzle and thermocouple for measuring gas temperature. Thermocouple in this photograph is 50 μm in diameter; the 13-μm-diameter thermocouple mainly used in the experiments is too thin to clearly show in the photograph. Coordinates $(X, Z)$ are defined with the nozzle tip center as the origin. (d) Temperature distribution in depth ($Z$) direction at $X$ = 5.5 mm. Red curve shows Gaussian fitting. Gray vertical lines show $1/e$-width, and red dashed line denotes $1/e$-width average temperature $T_{\mathrm{g}}^{1/\mathrm{e}}$.

### *3.2 Temperature-variable thermal radiation light sources*

Three custom-made planar blackbody light sources were used, depending on the required temperature range. Light source A (Fig. 4(a)) has an effective area of 130×130 mm square, a temperature control range of 5–105°C using a Peltier element, and temperature changing rates of 0.4°C/s for heating and −0.15°C/s for cooling. Light source B (Fig. 4(b)) has an effective diameter of 177 mm, a temperature control range from room temperature (25°C) to 430°C, and temperature rates of 0.7°C/s (heating) and −0.05°C/s (natural cooling). At high surface temperatures, buoyant flows of heated atmospheric $CO_2$ near the surface create flickering in the background image. In addition, when airflow hits the surface, temperature inhomogeneity on the surface exhibits nonuniformity in background radiation. To address these problems, the blackbody is encapsulated in a vacuum (< 50 Pa), and the infrared radiation is emitted through a double-sided, anti-reflection-coated Si window. Finally, light source C (Fig. 10(a)) has nearly the same appearance as light source B and is similarly vacuum-sealed, but it uses a water-cooled

Peltier element instead of a simple heater for its temperature control range from –20°C to 80°C and temperature changing rates of 0.11°C/s (heating) and −0.06°C/s (cooling). All radiation surfaces are coated with blackbody paint (0.986 emissivity in the $CO_2$ camera band). In this study, the surface temperature of these light sources serves as the absolute standard for gas temperature measurement. Therefore, the surface temperature was calibrated using a manufacturer-calibrated sensor. During the temperature scanning, the surface temperature is monitored in real time by a radiation thermometer with up to 10-ms time resolution, and it is converted to the equivalent blackbody temperature $T_s$, corrected for surface emissivity and Si window transmissivity. Radiation thermometers for light sources B and C are installed inside the vacuum chamber.

### *3.3 Gas supply system*

For proof-of-concept experiments, a system was built to supply $CO_2$ gas with controlled concentration, flow rate, and temperature. For concentration adjustment, $CO_2$ diluted with $N_2$ was used, since $N_2$ is infrared inactive and does not affect images. The required concentration and flow rate of $CO_2$ were generated by a mixed gas generator equipped with mass flow controllers (HORIBA, MU-3314). Two gas cylinders—a pure $CO_2$ cylinder and one with 4.0% $CO_2$ diluted in $N_2$—were used for supply. The temperature was controlled just before emission from the nozzle. Two custom-made gas temperature controllers were used according to the required temperature range. Gas temperature controller A (Fig. 4(a)) uses a Peltier element and is controllable in the 15–60°C range; gas temperature controller B (Fig. 4(b)) uses a heater and can control temperatures from room temperature (25°C) to 150°C. To release a sheet-like gas flow of constant thickness and column density, we used a flat stainless-steel nozzle with inner dimensions of 12.55×1.65 mm at the tip (Fig. 4(c)). The coordinate system ($X$, $Z$) was defined as shown in Fig. 4(c), with the nozzle center as the origin. Gas temperature control is not easy: Temperature rapidly converges to room temperature after ejection from the nozzle. Consequently, it is important to know the actual temperature of the emitted gas. However, in this study, the only available gas temperature sensors were thermocouples.

As discussed in Sec. 1, thermocouples are not considered reliable for gas temperature measurement due to thermal conduction through the probe and radiation from the probe's surface. Therefore, to maximize the reliability of the obtained values, we used an ultrafine thermocouple with minimized heat capacity. The temperature of the outflowing gas was monitored with a 13-μm-diameter thermocouple (Anbe SMT Co., KFT-13-200-100). An example of the temperature distribution measured with this thermocouple is shown in Fig. 4(d). This distribution is reasonably fit by a Gaussian function, confirming that the temperature distribution assumed in Sec. 2.4 was justified for this nozzle.

### *3.4 Experimental procedure*

A temperature-variable light source (Sec. 3.2) was placed behind the gas to be measured. The $CO_2$ imaging camera (Sec. 3.1), equipped with a 50-mm F2.5 lens, was properly positioned for observing the gas and the light source. For measuring gas having a known temperature, a gas supply system (Sec. 3.3) was used to emit $CO_2$ at the specified concentration $c$ and flow rate $Q$. For the demonstrations of engine exhaust and exhaled breath, the target gas contained $CO_2$. When the $CO_2$ in the gas was insufficient, pure $CO_2$ was inserted at the inlet to reach the required concentration. In all experiments, the $CO_2$ column density was within the effective range for the temperature measurement ($\lesssim$ 1000 ppm m) discussed in Sec. 2.4. Appropriate exposure time $t_{\mathrm{exp}}$ (2.5–40 ms) and frame rate $f$ (24.9–30 fps) were selected depending on the target, and the movie recording started simultaneously with the start of the temperature change. The data acquisition of the radiation thermometer of the light source was synchronized with that of the camera image, assigning a light source (background) temperature $T_s$ to each frame. The equivalent temperature $T_s^{eff}$ could be determined from $T_s$ by taking account of the transmissivity $\tau_b^{abs}$ of the background $CO_2$ based on Eq. (8). To determine the signal difference $\Delta I_R$ due to the presence of gas, a reference region on the image periphery unaffected by the gas was selected, and its average intensity was designated as $I_R(0)$. Occasional residual gas in the reference region was excluded using a median filter in the time domain. In the $\Delta I_R$ image, the background always exhibits a constant intensity, and the gas appears either brighter or darker relative to the background. More complex image processing was applied as needed, as discussed in the following sections. Image processing was performed by original software written in Python.

As discussed in the previous section, we had no other choice than to use a thermocouple to confirm the gas temperature in this study. Basically, the abovementioned 13-μm-diameter thermocouple was used, but when this one was bent by the flow, we adopted a 50-μm-diameter thermocouple (Anbe SMT, KFT-50-200-100).

Due to the uncertainty of the results from these thermocouples for gas temperature measurement, the absolute accuracy of the gas temperature determined by contrast reversal cannot be verified within this study. Confirmation of absolute accuracy must await comparison with other techniques in the future. Nevertheless, a thermocouple is a practically useful standard. Therefore, in this study, we compare the contrast reversal temperature with the gas temperature measured by the ultrafine thermocouple. Various other instruments were also used to clarify the experimental conditions (details described in Supplement 1, Sec. D).

## 4. Proof-of-concept experiments

### *4.1 Moderate-concentration $CO_2$ gas at room temperature*

To verify whether known temperatures could be reasonably measured, three proof-of-concept experiments were conducted. First, the simplest case—a gas at room temperature—was chosen, since there is no heat exchange with the surrounding environment and the gas temperature can be maintained uniformly. Under an ambient temperature of $T_a$ = 24.5°C, $CO_2$ gas with a concentration of $c$ = 10% and a flow rate of $Q$ = 1 L/min was ejected from gas temperature controller A, which was set at 25°C in the setup of Fig. 4(a). The column density $\zeta$ at the nozzle outlet was 165 ppm m, and the gas temperature was measured as $T_g$ = 24.8°C with a thermocouple. Using light source A, a movie was recorded during the process where $T_s^{eff}$ increased from 20°C to 30°C (assuming $\tau_b^{abs}$ = 0.874 for 64 ppm m) under conditions of $t_{exp}$ = 40 ms and $f$ = 24.9 fps. The results are shown in Fig. 5 and Visualization 1.

When $T_s^{eff} < T_g$, the gas appears bright (Fig. 5(a)), but when $T_s^{eff} > T_g$, the gas appears dark (Fig. 5(c)). In addition, there is a momentary transitional point when the gas becomes invisible (Fig. 5(b)). To discuss this process quantitatively, the line profile just behind the nozzle was examined. Figure 5(d) shows the transition of the line profile in the original $I_R$ image.

As predicted in Fig. 1(b), while $T_s^{eff}$ is low, the gas exhibits a positive contrast: As $T_s^{eff}$ increases, the contrast decreases, and at a certain $T_s^{eff}$, the image vanishes. At higher $T_s^{eff}$, the contrast reverses to negative. The baseline of the profile, except for the center region showing the ejected gas, presents the background intensity $I_R(0)$. However, it is not necessarily flat. This is mainly because the ejected $CO_2$ gas flowed in reverse due to changes in the environmental airflow. In the lower panel of Fig. 5(e), the change in the line profile of $\Delta I_R$ is illustrated as a color map. Here, the fluctuation of the baseline was removed by subtracting the original $I_R(0)$ fitted to a polynomial function.

The root mean square (RMS) of that profile is shown in the upper panel of Fig. 5(e). The contrast reversal temperature $T_s^{rev}$ (24.5°C) was obtained as the light source temperature at which the RMS reaches a minimum (i.e., the most featureless). This was regarded as the gas temperature $T_g^{1/e}$, and it agreed with the result obtained by the thermocouple $T_g$ within 0.3°C. The obtained $T_g^{1/e}$ value was closer to $T_a$ than $T_g$, which can be interpreted that the gas converged to the ambient temperature $T_a$ immediately after ejection.

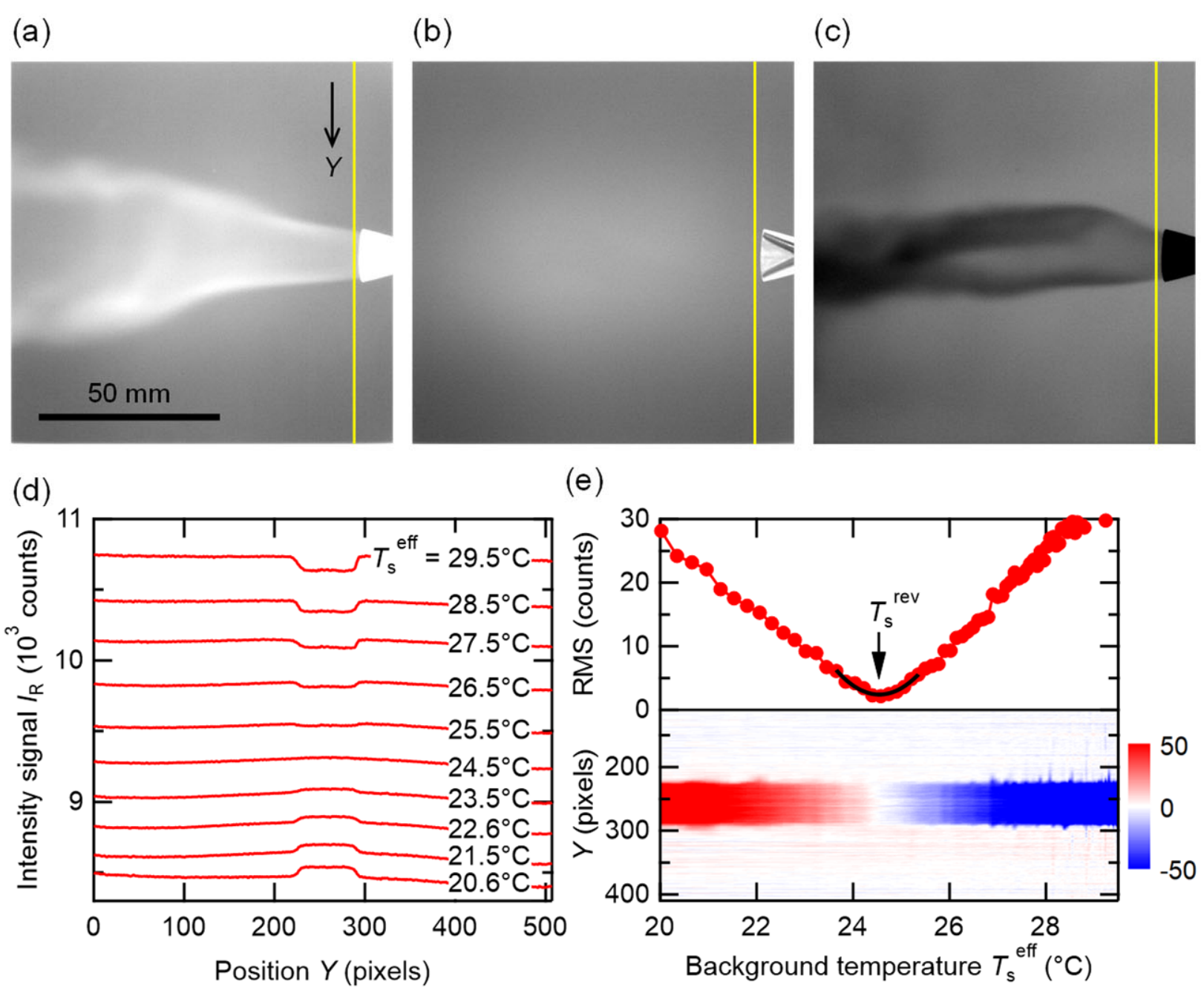

Fig. 5. Gas temperature determination by contrast reversal for a 10% $CO_2$ gas at room temperature. (a) Signal difference $\Delta I_R$ image at effective light source (background) temperature $T_s^{eff}$ = 20.6°C, (b) at 24.5°C, and (c) at 28.5°C, displayed with a common brightness scale (see Visualization 1). Yellow line shows position for intensity profile measurement. As in conventional image processing, positive $Y$ axis is defined downward. (d) Variation in intensity profile $I_R$ immediately past the nozzle as $T_s^{eff}$ increases. (e) Relationship of $\Delta I_R$ to $T_s^{eff}$, shown by a color map (bottom) and the transition of its RMS (top). Black curve shows the quadratic fitting for accurately determining the contrast reversal temperature $T_s^{rev}$.

### *4.2 Low-concentration $CO_2$ gas at room temperature: contrast enhancement*

Next, we confirmed that the contrast reversal method could also be applied to gases of lower concentrations. Under ambient temperature $T_a$ = 21.9°C, $CO_2$ gas at $c$ = 1% ($\zeta$ = 16.5 ppm m at nozzle outlet) and $Q$ = 1 L/min was emitted from gas temperature controller A set at 25°C. The setup was the same as that in Sec. 4.1, but this experiment was conducted on a different day at a lower room temperature. Regardless of the set value of the gas temperature controller (25°C), the gas temperature was determined as $T_g$ = 22.2°C with a thermocouple. The temperature of the nozzle itself was confirmed to be close to the ambient temperature ($T_a$) by thermocouple. Since the body of the temperature controller around the nozzle was exposed to the room environment, it is reasonable that the nozzle temperature deviated from the set value (25°C). Using light source A, a movie was recorded ($T_s^{eff}$ from 13°C to 27°C, $\tau_b^{abs}$ = 0.874 for 64 ppm m, $t_{exp}$ = 40 ms, and $f$ = 24.9 fps).

In this case, the line profile just behind the nozzle showed large fluctuations, and clear results like those in Fig. 5(e) could not be obtained. According to the $A_g$–ζ relationship for $T_g$ ~ 25°C (Fig. S1(a)), the absorptivity of $CO_2$ at the column density ζ = 16.5 ppm m is $A_g$ ~ 0.02. Considering the camera's dynamic range ($10^3$), visualizing this absorption is not difficult. The issue lies in the $CO_2$ originally present in the environment. The distance from the camera to the background light source was 0.9 m. Considering the atmospheric $CO_2$ concentration of 400 ppm, the column density in the atmosphere itself is 360 ppm m, 20 times higher than that from the nozzle (16.5 ppm m). Fluctuation due to air currents in the room obscured the intensity change caused by the gas emitted from the nozzle. Therefore, it is difficult to extract only the contribution of the target gas by observing solely the line profile at the nozzle exit.

However, the environmental $CO_2$ and the emitted gas from the nozzle behave differently. Therefore, we attempted to visualize only the gas from the nozzle based on differences in the image features. The gas from the nozzle changes more abruptly compared with the environmental $CO_2$. Therefore, it can be extracted by a time derivative. However, the simple difference from the previous frame introduces considerable noise. Therefore, to clarify the change in the image while reducing random noise, we evaluated the difference from the average of 30 surrounding frames (past 15 frames and next 14 frames; for about 1 second) at each moment. This is referred to as an enhanced difference image.

The results are shown in Fig. 6(a)–(c). At low $T_s^{eff}$, the gas motion is clearly visible (Fig. 6(a)). At a certain time, it disappears (Fig. 6(b)). Then, with a further increase in $T_s^{eff}$, the gas motion becomes visible again (Fig. 6(c)). As a quantitative indicator, the standard deviation of the enhanced difference image in the measurement box shown in the figure was investigated for each frame (Fig. 6(d)). The minimum value was observed at $T_s^{rev}$ = 22.2°C, indicating that the image is the most featureless at this $T_s^{eff}$. This is regarded as the gas temperature $T_g^{1/e}$, and it matched the result by thermocouple within 0.1°C. By making full use of 2D image information in this way, the temperature of the gas from the nozzle could be identified even in an environment with a 20-times-higher column density of $CO_2$ (360 ppm m vs 16.5 ppm m).

Visualization 2 shows the original (signal difference $\Delta I_R$) and enhanced difference images side by side. The environmental flow is also visible in the enhanced difference images, but the flow from the nozzle is dominant. Therefore, it was possible to identify the moment at which the gas of interest vanished, as shown in Fig. 6(d).

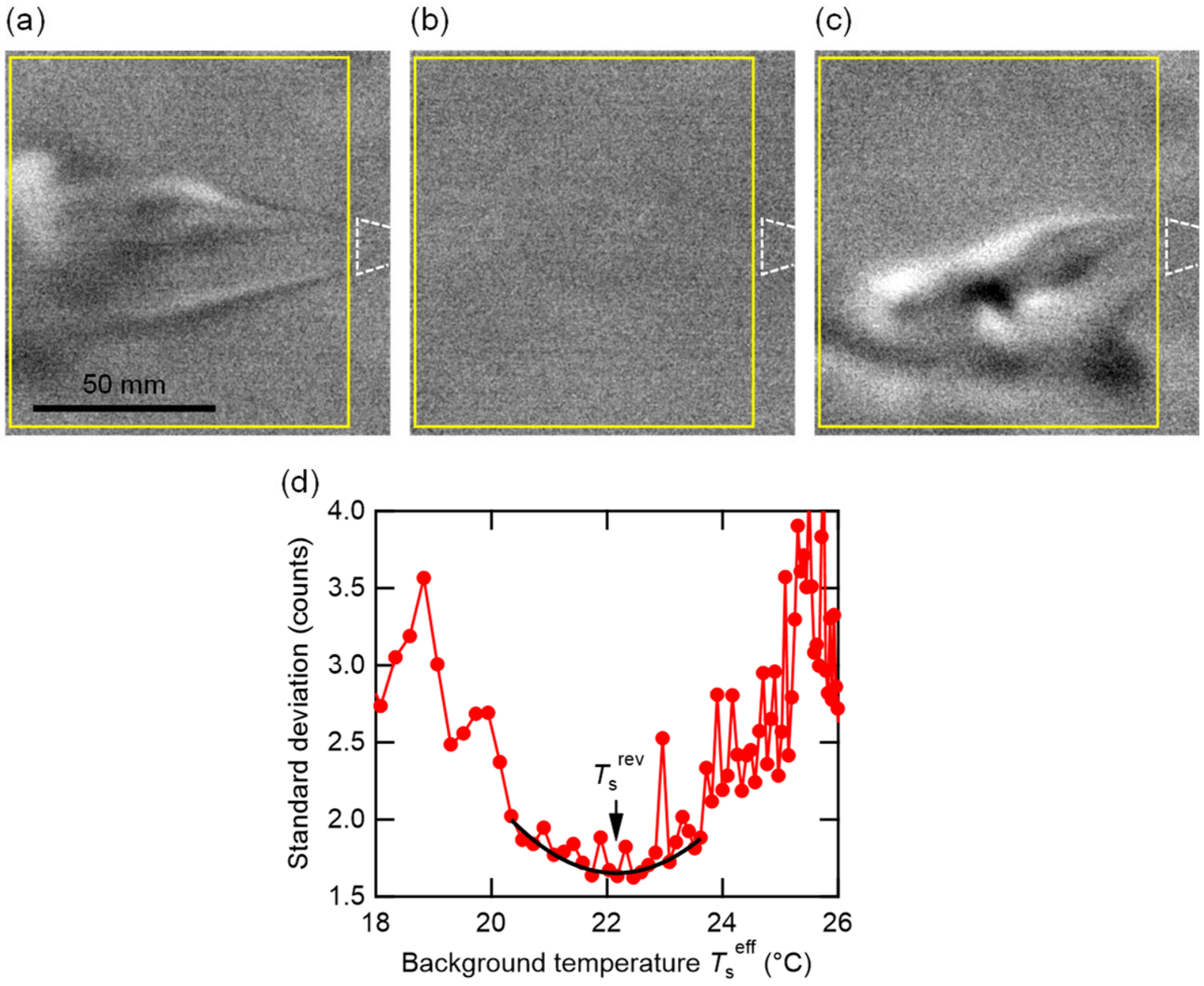

Fig. 6. Gas temperature determination by contrast reversal for a 1% $CO_2$ gas at room temperature. (a) Enhanced difference image of intensity $I_R$ at effective light source (background) temperature $T_s^{eff}$ = 18.1°C, (b) at 22.2°C, and (c) at 26.2°C, displayed with a common brightness scale (see Visualization 2). Yellow box indicates region for standard deviation measurement. Nozzle outline is shown by white dashed line. (d) Variation in standard deviation of enhanced difference image in measurement region as $T_s^{eff}$ increases.

### *4.3 Hot $CO_2$ gas with moderate concentration: 2D mapping*

In the final proof-of-concept experiment, it was revealed how contrast reversal occurs in gas at a substantially higher temperature than room temperature. Under ambient temperature $T_a$ = 24.6°C, $CO_2$ gas with $c$ = 10% ($\zeta$ = 165 ppm m (100°C) at the nozzle outlet) and $Q$ = 3 L/min was ejected from gas temperature controller B set at 100°C. The gas temperature was measured as $T_g$ = 106.6°C by thermocouple.

Using the setup in Fig. 4(b) and light source B, a movie was recorded ($T_s^{eff}$ from 27°C to 111°C, $\tau_b^{abs}$ = 0.880 for 60 ppm m, $t_{exp}$ = 5 ms, and $f$ = 30 fps). The gas temperature was continuously connected with surrounding $T_a$ and had a distribution in the $Z$ (depth) direction. Figure 4(d) shows the temperature distribution at $X$ = 5.5 mm during this experiment. The gas temperature also converges to $T_a$ as it moves farther away from the outlet in the +$X$ direction. Representative $\Delta I_R$ images at three different $T_s^{eff}$ values are shown in Fig. 7(a)–(c) and Visualization 3. Nevertheless, different from the previous cases, we could not find any moment when the gas completely disappeared.

The variation in the average intensity of the square region just past the nozzle (shown as boxes in Fig. 7(a)–(c)) is shown in Fig. 7(d). To exclude rapid fluctuations in gas, a moving average of $N$ = 30 points (for 1 s) is shown, but abrupt negative spikes are still remarkable. This results from the backward flow of dense $CO_2$ into the measurement region due to changes in airflow in the environment; here, the cooled, previously ejected $CO_2$ appears dark, so these spikes always appear as negative. On the other hand, in regions far from the nozzle, the gas flutters randomly; the presence and absence of the gas are randomly repeated. These stochastic fluctuations can be removed with a median filter of suitable length $N$ in the time domain, which gives the most representative value among $N$ continuous frames by excluding exceptional frames. Here, $N$ = 5 was chosen. Smooth trends for each pixel were obtained by fitting a polynomial function. The $T_s^{\mathrm{eff}}$ when this trend crosses $\Delta I_{\mathrm{R}} = 0$ was determined to be $T_s^{\mathrm{rev}}$ for each local region. In areas without the gas, the background $I_{\mathrm{R}}(0)$ is visible most of the time, so $\Delta I_{\mathrm{R}}$ is always ≈ 0 and the correlated trend with $T_s^{\mathrm{eff}}$ as in Fig. 7(d) is not observed. Therefore, as a first step, abnormal points where the $\Delta I_{\mathrm{R}}$ trend does not cross 0 and areas of the nozzle itself or those outside the light source window were excluded. Furthermore, we excluded points where the width $W$ of the change (Fig. 7(d)) over the entire experiment duration was lower than a certain threshold. The resulting temperature distribution for each valid pixel is shown in Fig. 7(e). In this way, the representative gas temperature at each pixel could be determined, even for dynamically fluttering gas.

Two representative $T_s^{\mathrm{rev}}$ profiles at $X$ = 0.5 mm and 5.5 mm at the nozzle center extracted from Fig. 7(e) are shown in Fig. 7(f) and, in Table 1, compared with the temperature profiles in the $Z$ direction measured by thermocouple. For the thermocouple measurements, both $T_g^{\mathrm{peak}}$ and $T_g^{1/e}$ are displayed. $T_s^{\mathrm{rev}}$ is the value at the position of the white circle (center line) in Fig. 7(f). While the temperature $T_s^{\mathrm{rev}}$ obtained from the contrast reversal was closer to $T_g^{1/e}$ than to $T_g^{\mathrm{peak}}$ as expected, a discrepancy of 5–7°C was found between $T_s^{\mathrm{rev}}$ and $T_g^{1/e}$. Nevertheless, considering that the temperature of a dynamically fluctuating gas was evaluated remotely and noninvasively, the advantage of the contrast reversal method proved to be sufficient. There could also be underestimation in the results of the thermocouple due to heat escape. Figure 7(e) also illustrates how high-temperature gas at ~ 100°C ejected into a room-temperature environment rapidly cools down; here, the gas temperature is below 40°C after traveling 50 mm.

This procedure for measuring temperature distribution was adopted in all of the following cases as well. Only the appropriate value of $N$ for the median filter was adjusted for each case.

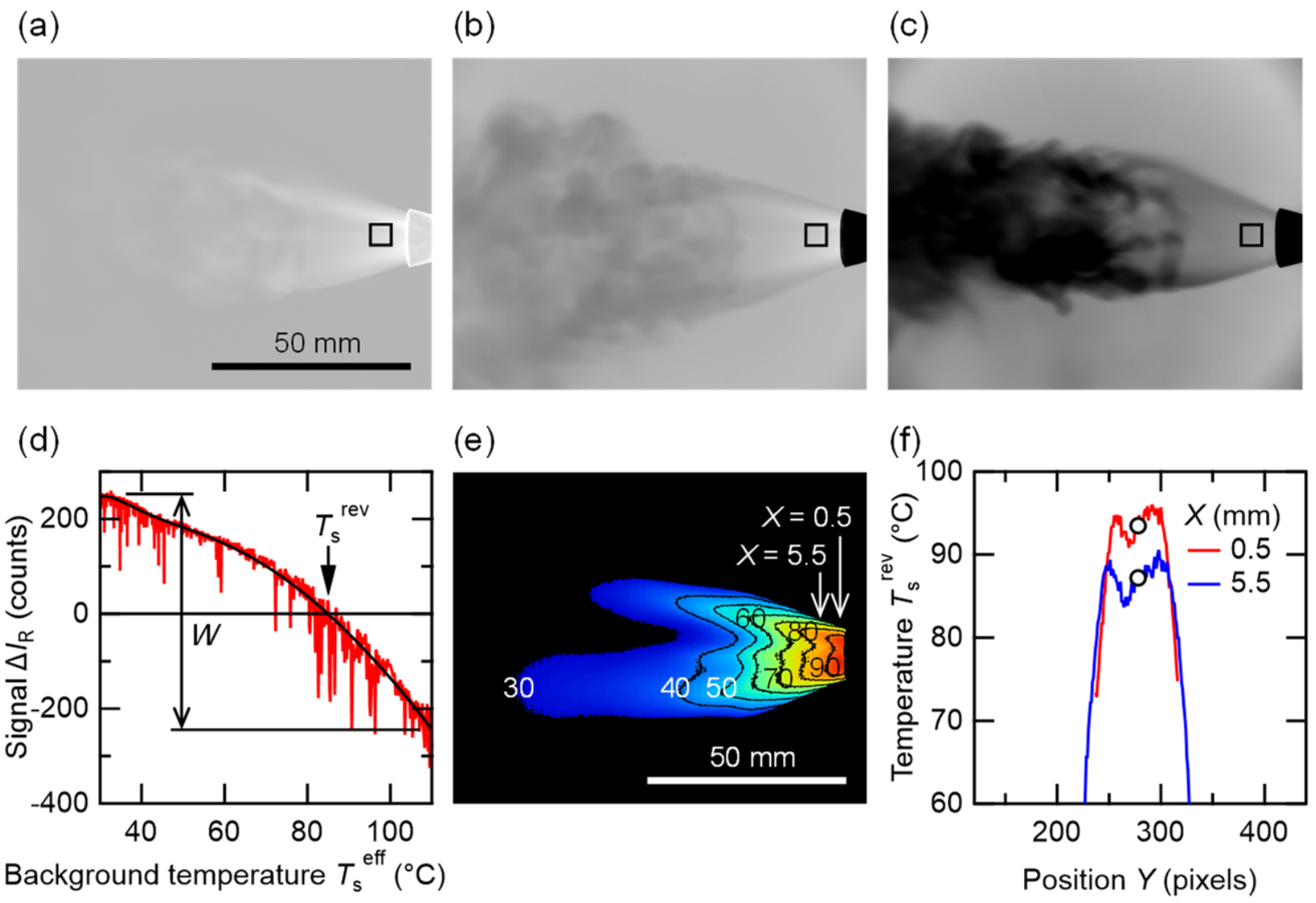

Fig. 7. Gas temperature determination by contrast reversal for a 10% $CO_2$ gas at ~ 100°C. (a) Signal difference $\Delta I_{\mathrm{R}}$ image at the effective light source (background) temperature $T_s^{\mathrm{eff}}$ = 30°C, (b) at 70°C, and (c) at 110°C, displayed with a common brightness scale (see Visualization 3). (d) Variation in average $\Delta I_{\mathrm{R}}$ of the square region just past the nozzle shown in (a)–(c) as $T_s^{\mathrm{eff}}$

increases. (e) 2D temperature distribution by obtaining the contrast reversal temperature $T_s^{rev}$ (°C) at each pixel. Invalid region is shown by black. (f) Temperature profile at representative positions in (e).

**Table 1. Representative temperatures at different $X$ positions for a 10% $CO_2$ gas at ~ 100°C. Comparison of the peak gas temperature $T_g^{peak}$ and the $1/e$-width average value $T_g^{1/e}$, both determined from $Z$-direction temperature distribution measured by thermocouples, and contrast reversal temperature $T_s^{rev}$, determined from signal difference $\Delta I_R$ images.**

| Position $X$ (mm) | Thermocouple | | $CO_2$ image |
|---|---|---|---|
| | $T_g^{peak}$ (°C) | $T_g^{1/e}$ (°C) | $T_s^{rev}$ (°C) |
| 0.5 | 106.6 | 88.2 | 93.5 |
| 5.5 | 97.7 | 80.3 | 87.2 |

## 5. Application experiments

### 5.1 Hair dryer

The method described in the previous section was applied to various forms of gases at unknown temperatures. As an example of continuous high-temperature gas, we measured the temperature of the air blown from a hair dryer (Panasonic EH534, power: 900 W). With our $CO_2$ camera, it was possible to visualize the air emitted from the hair dryer as is, since the atmospheric $CO_2$ at 400 ppm (0.04%) is heated and appears bright. However, the contrast was insufficient to determine the gas temperature. Therefore, pure $CO_2$ gas as a tracer was mixed into the intake air. By mixing $CO_2$ gas at 23.5 L/min, the emitted $CO_2$ concentration increased to $c$ = 3.8%. From the $CO_2$ flow rate and concentration, the total flow rate of the hair dryer is estimated to be $Q$ = 620 L/min. The outlet has an opening of 21 mm in the depth ($Z$) direction, and the column density is estimated at $\zeta$ = 800 ppm m (100°C). The setup is shown in Fig. 8(a). Using light source B, a movie was recorded ($T_a$ = 25.8°C, $T_s^{eff}$ = 31–156°C, $\tau_b^{abs}$ = 0.880 for 60 ppm m, $t_{exp}$ = 2.5 ms, and $f$ = 30 fps).

Typical images at low and high $T_s^{eff}$ are presented in Fig. 8(b) and (c), respectively, as well as in Visualization 4. As described in Sec. 4.3, the temperature distribution for each pixel shown in Fig. 8(d) was determined ($N$ = 5 for median filter). The maximum $T_s^{rev}$ at the outlet was 107°C; the temperature was distributed asymmetrically in the vertical direction, and the upper side was hotter. The temperature measured with a 50-μm thermocouple exhibited the maximum temperature of 140°C near the upper end, and $T_g^{1/e}$ was estimated at 111°C. This result is consistent with the contrast reversal method (107°C). From the obtained temperature distribution, at a distance of 100 mm, reflecting a typical position for drying hair, the temperature was estimated at $T_g^{1/e}$ ~ 65°C and $T_g^{peak}$ ~ 80°C.

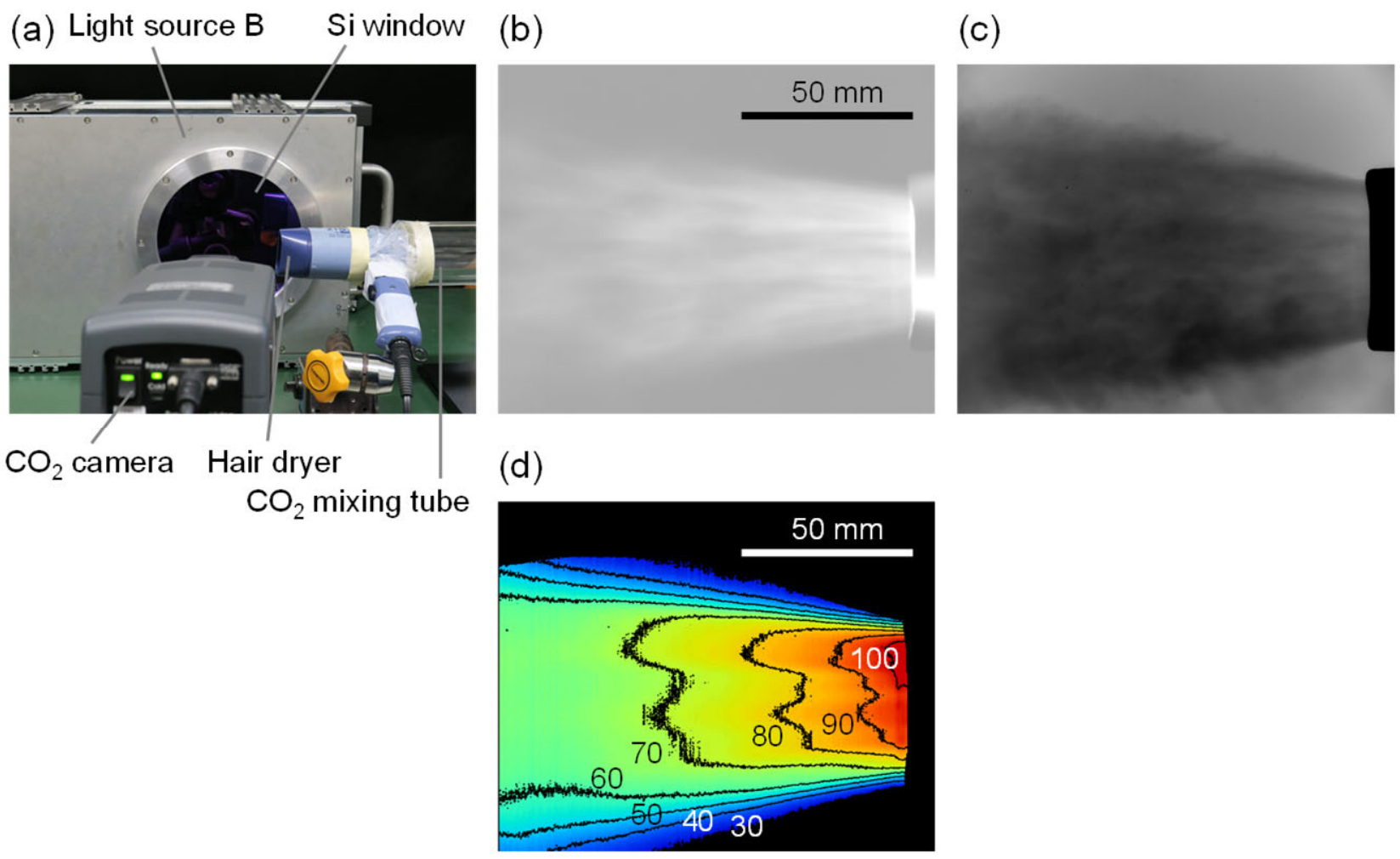

Fig. 8. Temperature measurement of air from a hair dryer. (a) Measurement setup. A cylinder was fixed at the intake of the dryer, and pure $CO_2$ gas was introduced from the side. (b) Signal difference $\Delta I_R$ image at effective light source (background) temperature $T_s^{eff}$ = 35°C and (c) at 135°C, displayed with a common brightness scale (see Visualization 4). (d) 2D distribution of the contrast reversal temperature $T_s^{rev}$ (°C).

### *5.2 Exhaust gas of a diesel engine: quick intermittent emission*

Next, we measured the temperature of exhaust gases from a diesel engine vehicle. This differs from previous cases in that the gas is ejected intermittently. Additionally, the camera and light source were taken outside the laboratory for measurement, making this experiment meaningful as a practical demonstration.

We used an agricultural tractor, the John Deere 1750 (2.940 L, 3 cylinders). The exhaust pipe was modified to protrude to the side, making observation more convenient. Below the exhaust pipe with an elliptical cross section (inner diameter: 50 mm horizontally, 45 mm vertically), light source B was placed facing upward, and the $CO_2$ camera was set above facing downward (Fig. 9(a)).

In an environment at $T_a$ = 12.8°C, the engine speed was maintained at 900 rpm (idling). The average exhaust flow was estimated to be $Q$ = 22 L/s. The exhaust gas sensor (thermocouple) gave $T_g^{peak}$ = 80°C ($T_g^{1/e}$ = 66°C) and $c$ = 1.9% ($\zeta$ = 855 ppm m (80°C)) at the center of the exhaust pipe's end. A movie was recorded during the rise of $T_s^{eff}$ from 32°C to 133°C ($\tau_b^{abs}$ = 0.853 for 80 ppm m, $t_{exp}$ = 3.5 ms, and $f$ = 30 fps). Typical $\Delta I_R$ images at low and high $T_s^{eff}$ are shown in Fig. 9(b) and (c), respectively, and in Visualization 5.

In Supplement 1, Sec. E and Fig. S4, we discuss how the temperature of such intermittently emitted gas is measured by the contrast reversal method. For simplicity, it was assumed that gas with constant $CO_2$ concentration and temperature was emitted as rectangular pulses. A schematic of contrast reversal for this case is shown in Fig. 9(d). Initially, when $T_s^{eff}$ is low, the gas looks bright in most frames but occasional frames without a gas image give negative spikes. When $T_s^{eff}$ is high, the gas is mostly dark, but some frames without a gas yields positive spikes. Therefore, even for intermittently emitted gases, by excluding frames where the gas was absent, the gas temperature can be determined using the contrast reversal method.

More specifically, a 4-stroke, 3-cylinder engine rotating at 900 rpm exhausts gas 22.5 times per second [42–44]. As discussed in detail in Supplement 1, Sec. E, by assuming the exhaust valve opening angle as 180 degrees, phases of producing emission for 33.3 ms and halting it for 11.1 ms were alternately repeated. Since $t_{\mathrm{exp}}$ is sufficiently shorter than these periods, some frames did not capture exhaust gases (Fig. S5). This is because the engine and the frame rate were not synchronized, and the frame rate was not fast enough to track all engine processes. Consequently, frames without gas appeared randomly with a probability of 25%.

Again, a temporal median filter is effective here. By choosing $N$ = 31 (about 1 s), the influence of gas-free frames could be excluded. By fitting the median-filtered $\Delta I_{\mathrm{R}}$–$T_{\mathrm{s}}^{\mathrm{eff}}$ for each pixel to a polynomial function to obtain a smoothed trend, the $T_{\mathrm{s}}^{\mathrm{eff}}$ at which $\Delta I_{\mathrm{R}}$ = 0 was determined as $T_{\mathrm{s}}^{\mathrm{rev}}$. As in Sec. 4.3, the temperature distribution for each pixel was determined as shown in Fig. 9(e).

At the center just after the outlet, $T_{\mathrm{s}}^{\mathrm{rev}}$ was 63.5°C, consistent with the $T_{\mathrm{g}}^{1/\mathrm{e}}$ (66°C) estimated by the exhaust gas sensor. A notable feature of this case is that the maximum temperature is not located immediately after the outlet. The maximum temperature point appears about 30 mm away from the outlet. This feature was common to other engine speeds (not shown). As shown in Fig. S5 and Visualization 5, gas released from the pipe does not flow away smoothly at a constant speed but forms mushroom-shaped vortices and exhibits halted motion around this position. As a result, the high-temperature gas at the center flows outward by the vortex and stagnates there, causing the maximum temperature region to appear slightly ahead of the outlet. However, intensive discussion on combustion engineering is out of the scope of this work. In addition, Fig. 9(e) shows that at 90 mm from the outlet, the gas cools to $T_{\mathrm{g}}^{1/\mathrm{e}} \approx$ 40°C ($T_{\mathrm{g}}^{\mathrm{peak}} \approx$ 45°C).

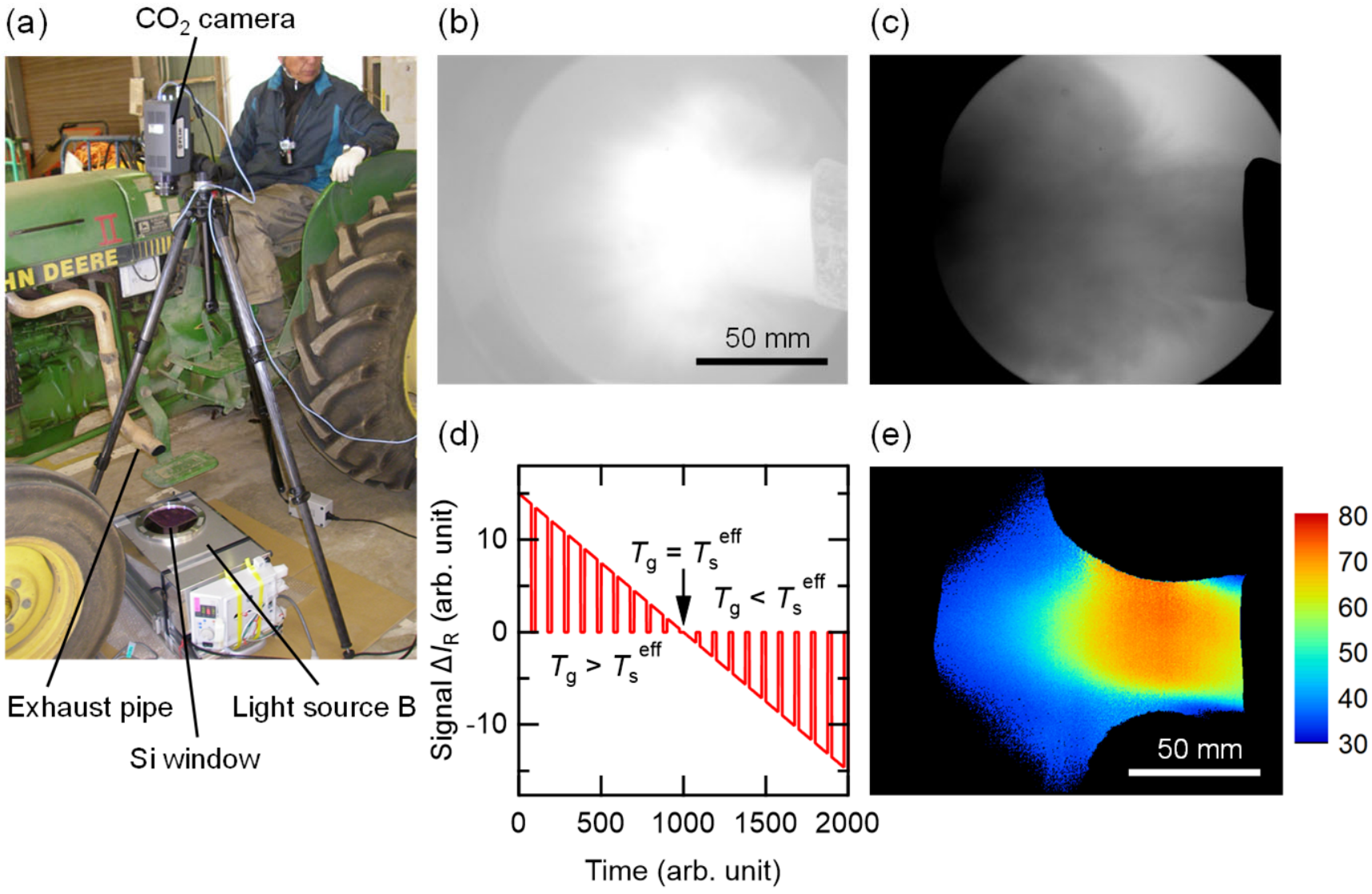

Fig. 9. Temperature measurement of exhaust gas from a diesel engine. (a) Measurement setup. (b) Signal difference $\Delta I_{\mathrm{R}}$ image at effective light source (background) temperature $T_{\mathrm{s}}^{\mathrm{eff}}$ = 31.8°C and (c) at 107.8°C, displayed with a common brightness scale (see Visualization 5). (d) Variation in $\Delta I_{\mathrm{R}}$ with respect to $T_{\mathrm{s}}^{\mathrm{eff}}$ based on a model of intermittently emitted gas. (e) 2D distribution of the contrast reversal temperature $T_{\mathrm{s}}^{\mathrm{rev}}$ (°C).

### *5.3 Human breath: slow intermittent emission*

As a relatively slow intermittent case, human breath was also measured. The subject was a 62-year-old male, with a body temperature of 36.0°C. The $CO_2$ concentration in the subject's breath was $c$ = 4.5% (end-tidal carbon dioxide ($EtCO_2$): 34.5 mmHg by a capnometer), and total exhaled air was $Q$ = 6.1 L/min (minute volume (MV): 12.2 L/min by a spirometer). From his mouth and nostrils dimensions in the depth ($Z$) direction, typical column density was estimated at ζ = 900 ppm m (35°C) for the mouth and 1125 ppm m (35°C) for the nose (both nostrils).

The setup is illustrated in Fig. 10(a). Using light source C, in an environment of $T_a$ = 25.2°C, a movie was recorded as $T_s^{\text{eff}}$ dropped from 38°C to 28°C ($\tau_b^{abs}$ = 0.829 for 100 ppm m, $t_{\text{exp}}$ = 40 ms, and $f$ = 24.9 fps). To achieve uniform temperature distribution on the light source surface, we searched for $T_s^{\text{rev}}$ while lowering $T_s^{\text{eff}}$. The subject placed his chin on a fixed rod to keep the head position constant.

$\Delta I_R$ images at three representative $T_s^{\text{eff}}$ values are shown in Fig. 10(b)–(d) and in Visualization 6. The subject breathed through both the nose and mouth. At first (high $T_s^{\text{eff}}$), the exhaled breath from both the mouth and nose appeared dark (Fig. 10(b)). At the end (low $T_s^{\text{eff}}$), both appeared bright (Fig. 10(d)). At an intermediate state, however, there was a moment where the exhaled breath from the mouth appeared bright while that from the nose was dark (Fig. 10(c)). This indicates that the exhaled breath temperature from the mouth differs from that from the nose.

As in Sec. 5.2, this is an intermittent gas, but the cycle is sufficiently slower than the frame rate; therefore, emission and halted phases are clearly distinguished (Fig. S4(c)). There were 21 breaths during the recording, and these were called B1, B2, ..., B21. Measurement regions of interest (ROIs) were set approximately 3 mm from both the mouth and nose (see boxes in Fig. 10(c)), and the average intensity $I_R$ within each ROI was evaluated. ROI sizes were 1.7×0.85 mm for the mouth and 2.7×2.7 mm for the nose.

Overall, stagnated gas originated from the previous exhalation tended to disturb the measurement. Each ROI was minimized to avoid such disturbances. Figure 10(e) shows the raw intensity signal $I_R$ of the mouth from B1 to B21. The times of the start and end of each exhalation were determined from the change in the image around the ROI for the mouth; these are shown as vertical gray lines. The intensity value where the $I_R$ curve intersects the gray line is assumed to indicate the background intensity. A smooth polynomial fit through these points (blue curve in Fig. 10(e)) is regarded as $I_R(0)$. Within the raw intensity $I_R$, some parts were excluded as invalid (black curve) due to the overlap with the previous exhalation. The remaining part (red curve) was valid, and its difference from $I_R(0)$, $\Delta I_R$ was used for temperature determination. A similar procedure was done for the nose (not shown).

Figure 10(f) shows the $\Delta I_R$–$T_s^{\text{eff}}$ relation for exhaled breath from the mouth and nose. The average $\Delta I_R$ for each breath (B1–B21), from start to end of the exhalation, is plotted as black dots. The intersection of the quadratic fit to these points with $\Delta I_R$ = 0 was defined as the contrast reversal temperature $T_s^{\text{rev}}$, which was found to be 33.7°C for the mouth and 32.8°C for the nose. The peak temperatures $T_g^{\text{peak}}$ are estimated to be 36.5°C for the mouth and 35.3°C for the nose, which are both close to body temperature.

Breath temperature was also measured with a 50-μm thermocouple (Fig. S6, see Supplement 1, Sec. F for details). Those temperatures were lower than those derived from $CO_2$ imaging by ~ 2°C: 31.9°C for the mouth and 31.0°C for the nose. These values are consistent with past reports [45]. The discrepancy of ~ 2°C can be attributed to the difference in measuring point. The thermocouple was positioned 10 mm from the mouth or nose, since it was difficult to position fragile probes any closer to a moving human subject. In the $CO_2$ image in Fig. 10(c), exhaled bright air from the mouth changes to dark just 10–20 mm beyond the mouth, supporting the expectation of rapid temperature drop over a short distance. Moreover, thermocouple measurements may be underestimated due to heat loss. However, the finding that mouth breath is ~ 1°C warmer than nasal breath is consistent for both imaging and thermocouple measurements.

Given that nose surface temperature is particularly low in the facial region [46–48], it is reasonable that breath exhaled through the nasal cavity is cooler than that through the mouth. Additionally, as seen in Supplement 1, Sec. F (Figs. S7–S8), interesting findings could be derived from our results, but further medical discussion is beyond the scope of this work. In this study, it was necessary to use complex manual processing to obtain $T_s^{rev}$ from breath images; this should be automated in the future. Nevertheless, the capability to successfully determine breath temperature based on contrast reversal is a significant achievement.

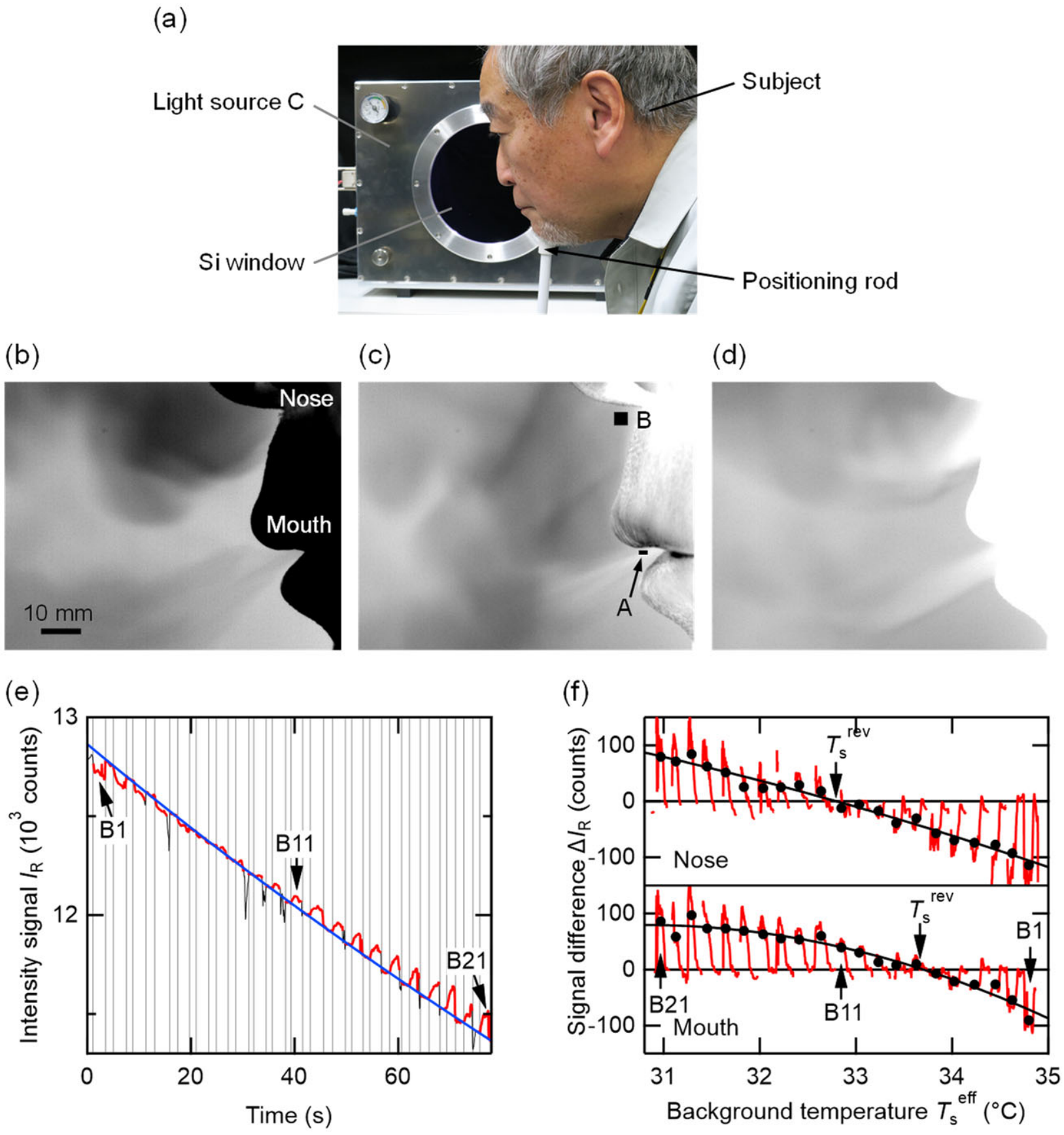

Fig. 10. Temperature measurement of human breath. (a) Measurement setup. (b) Signal difference $\Delta I_R$ image at effective light source (background) temperature $T_s^{eff}$ = 34.8°C, (c) at 33.0°C, and (d) at 30.9°C, displayed with a common brightness scale (see Visualization 6). Black boxes in (c) indicate measurement ROIs for the mouth (A) and nose (B). (e) Raw intensity signal $I_R$ of the mouth versus time. Vertical gray lines indicate timing of exhalation start and end. Invalid regions due to inflow of previously exhaled, cooled breath into ROI are shown by black curves. Red curves indicate valid regions. Numbers at representative breaths are denoted. Blue curve represents background intensity signal $I_R(0)$. (f) Relationship of valid $\Delta I_R$ with $T_s^{eff}$ for nose (top) and mouth (bottom). Black dots represent average $\Delta I_R$ for each breath (B1–B21) plotted at the midpoint between start and end of exhalation. Their quadratic fittings are shown by black curves; their intersection with $\Delta I_R = 0$ indicates the contrast reversal temperature $T_s^{rev}$.

## 6. Discussion

### 6.1 Temperature measurement accuracy

In previous temperature measurement based on the OGI technique [17,30], the temperature distribution of the gas in the depth direction was neglected and treated as a line-of-sight average or path-integrated temperature. In contrast, this study explicitly considered the temperature distribution in the depth direction and showed that the proposed method provides a temperature $T_{\mathrm{s}}^{\mathrm{rev}}$ close to the $1/e$-width average temperature $T_{\mathrm{g}}^{1/\mathrm{e}}$ for a gas with temperature and concentration distributed according to a Gaussian function.

Within this study, we could not confirm the absolute accuracy of the gas temperature obtained by the contrast reversal method. In actual experiments, at around 100°C (Sec. 4.3), there was a discrepancy of up to 7°C between $T_{\mathrm{s}}^{\mathrm{rev}}$ and $T_{\mathrm{g}}^{1/\mathrm{e}}$ measured by the thermocouple. However, we cannot discuss true accuracy as long as it relies on comparison with thermocouple measurements, which are not necessarily reliable. While the contrast reversal method was applied to targets such as engine exhaust and human breath in the application experiments, it is not clear whether their actual temperature distributions can be approximated by a Gaussian function. Further discussion on accuracy would require direct comparison of measurements from the contrast reversal method and advanced three-dimensional laser spectroscopy on the same targets.

### 6.2 Required $CO_2$ concentration for contrast reversal method

In the hair dryer example in Sec. 5.1, application of the contrast reversal method was judged to be difficult without the addition of $CO_2$. The original column density was estimated to be $\zeta = 8$ ppm m (100°C). However, in Sec. 4.2, contrast reversal could be applied to a similarly low $\zeta$ of 16.5 ppm m (25°C) by making full use of the overall image fluctuations. From these results, we can conclude that this method is applicable to $\zeta \gtrsim$ 10–15 ppm m. Based on the $A_{\mathrm{g}}$–$\zeta$ relationship (Fig. S1(a)), for $\zeta = 8$ ppm m (100°C) and 16.5 ppm m (25°C), $A_{\mathrm{g}} = 0.018$ and 0.031, respectively. Therefore, when absorptivity is ~ 3%, the contrast reversal can be applied. On the other hand, Fig. 3(c) suggests that when $\zeta \gtrsim 1000$ ppm m, the discrepancy between $T_{\mathrm{g}}^{1/\mathrm{e}}$ and $T_{\mathrm{s}}^{\mathrm{rev}}$ becomes substantial, leading to reduced accuracy of temperature measurement.

The contrast reversal method can also be applied to the measurement of other gases, with an OGI camera tailored to each target gas. For a variety of gases besides $CO_2$, this method should be applicable at even lower concentrations, since the presence of atmospheric $CO_2$ fluctuations limited the minimum application range in the case of measuring $CO_2$.

### 6.3 Comparison with classical line reversal method

The line reversal method in the 4-μm band for temperature measurement of $CO_2$-containing gases (flames) was demonstrated nearly 100 years ago by Henning *et al*. They placed an intensity-variable light source behind the flame and monitored the intensity at 4.4 μm using a monochromator and a thermocouple. Here, we employed a narrowband infrared camera. This led to four essential innovations. First, the use of high-sensitivity cameras permitted temperature measurement of gases of moderate temperatures, including those around room temperature. Traditionally, application of the line reversal method was limited to combustion gases of ~ 1000 K or higher. However, due to the complexity of combustion processes, the interpretation was not straightforward. There has been a debate over the presence of temperature error due to the chemical reactions [1,10,13]. In contrast, the gases in this study (~ 100°C) are simply thermodynamically hot, ensuring reliability of the determined temperature. Second, by leveraging established image processing techniques, rich image information could be fully exploited to enhance the accuracy of temperature determination, even from images with poor contrast. Third, 2D temperature distribution measurements were achieved. Finally, use of high-speed cameras made it possible to apply the method to dynamic targets such as internal combustion engines and human respiration.

### 6.4 Significance of current study

The ability to measure 2D temperature distributions simply by capturing images with a camera is expected to find many applications. The gas from the flat nozzle in Sec. 4.3 rapidly cooled, but in Sec. 5.1, the gas from a hair dryer was delivered efficiently over a greater distance. This confirmed that the hair dryer was well-engineered to satisfy its functional requirements. This technique is expected to play an important role in the development of such fluidic devices. Section 5.2 suggests the technique's usefulness for the development of internal combustion engines. Section 5.3 demonstrated that this approach can also provide valuable insights in the medical field. It not only clearly illustrated the temperature differences in exhaled breath but also clarified the fast and complex dynamics of human breathing.

Another important objective is the quantitative measurement of $CO_2$ emissions. Equation (7) shows that $A_\mathrm{g}$ can be obtained from the intensity difference $\Delta I_\mathrm{R}$ if the gas temperature is known. Once $A_\mathrm{g}$ is determined, the column density and thus the total amount of gas can be estimated [16,32]. Therefore, gas temperature measurement is the first step to remotely quantifying $CO_2$ emission. Toward realizing a low-carbon society, $CO_2$ emission measurement based on gas imaging will make significant contributions to the detection of $CO_2$ leaks and estimation of emission volumes.

### 6.5 Required mid-infrared devices

A current challenge in applying this method is the time required. The temperature scanning for contrast reversal typically takes several minutes due to the slow response of thermal radiation light sources. Recently, LEDs in this wavelength range have become available; however, their operation is limited to pulsed mode and their intensity is insufficient. Consequently, implementation of electroluminescent planar light sources is highly anticipated [49]. This would enable not only high-speed scanning, but also eliminate the need for vacuum encapsulation, allowing a simpler system for gas temperature measurement compared with the present work.

In addition, OGI cameras usually require cooling to below 80 K, and thus they are expensive. However, the development of narrowband infrared detectors with higher operation temperature is making significant progress, based on optimally engineered quantum wells [50–52]. Therefore, the development of inexpensive OGI cameras is also anticipated.

### 6.6 Comparison with conventional technologies and future challenges

Among various laser spectroscopy techniques for temperature measurement, tunable diode laser absorption spectroscopy (TDLAS) is particularly well developed. Through comparison with TDLAS, we would like to clarify the future challenges of the method proposed in this study.

The essential difference between TDLAS and the contrast reversal method (or OGI) derives from their underlying technologies: narrow-linewidth lasers or image sensors. TDLAS is an active technique based on near-infrared semiconductor laser technology [53]. The laser light has a linewidth much narrower than individual absorption lines of the gas, and its intensity is overwhelmingly higher than the thermal radiation from the background or gas. Therefore, discussion on a specific transition of a particular gas species is theoretically straightforward, and high-speed measurements are also possible. While the temperature determined with TDLAS was initially limited to line-of-sight average temperature [54,55], nowadays tomographic determination of temperatures or gas concentrations at specific points in space has been achieved. For example, tomographic imaging of aircraft engine exhaust [56] and combustion diagnosis of temperature and multiple gas concentrations have been demonstrated [57].

On the other hand, OGI is a fundamentally passive technique based on mid-infrared image sensor technology. The intensity observed is comparable with the thermal radiation of room-temperature objects or gases, and there are even cases where the radiation from the gas exceeds that from the light source (background). In addition, the linewidth of interest is broad, observing the integration of many absorption lines. In this study, we discussed nondispersive OGI, i.e., OGI that does not use precise spectroscopic elements. Compared with dispersive OGI, this has the advantage of being able to dynamically track the movement of specific gases. However, compared with state-of-the-art laser spectroscopy techniques like TDLAS, the supporting fundamental technologies are still immature. As discussed in Sec. 6.5, mid-infrared devices have not advanced as much as near-infrared devices. Nevertheless, with the recent spread of infrared cameras and the growing attention towards greenhouse gas emissions, OGI-based technology is expected to attract greater interest in the future.

For the establishment of the contrast reversal method, it is necessary to verify the temperature measurement accuracy by comparison with established methods such as TDLAS, to clarify the contrast reversal temperature for arbitrary temperature distributions, and to develop mid-infrared devices for achieving high-speed temperature (intensity) scanning, system simplification, and broadened range of applicable gases. Attempts at three-dimensional measurements using multiple OGI cameras have already begun for concentration measurements [58,59]. As discussed in Sec. 6.4, temperature measurement based on the contrast reversal would be particularly useful as a step toward quantitative measurement of gas emission based on OGI.

## 7. Summary

This study demonstrated a noninvasive method for measuring the temperature of $CO_2$-containing gases released into free space, based on the image contrast reversal produced by a temperature-variable light source and a $CO_2$ imaging narrowband mid-infrared camera. By sweeping the temperature of the light source placed behind the gas, the gas temperature is determined as the temperature of the light source when the gas body's image contrast reverses. For a gas plume with Gaussian-distributed temperature and concentration, the contrast reversal temperature yields a value close to the $1/e$-width average temperature. The method was applied to various $CO_2$ gases with temperatures ranging from 25°C to 100°C and concentrations (column densities) ranging from 15 to 1000 ppm m, achieving gas temperature determinations consistent with the results by the thermocouple within a range of 0.1–7°C. The measurement of 2D temperature distributions and its application to dynamically emitted gases are also possible. Moreover, this method can be applied to gases lacking sufficient $CO_2$ by simply adding $CO_2$. The minimum required $CO_2$ concentration was reduced by leveraging established image processing technologies. Applications to various target sources, including outdoor combustion engines and humans, were also demonstrated. The proposed method is also expected to be effective for quantitative $CO_2$ emission measurements.

**Funding.** National Institute for Materials Science (Sensors and Actuators Research Project); Cabinet Office (PRISM); Ichimura Foundation of New Technology; Steel Foundation for Environmental Protection Technology; Japan Society for the Promotion of Science (JP22K18990, JP23K26576, JP23H01883); New Energy and Industrial Technology Development Organization (JPNP14004).

**Acknowledgments.** The authors are thankful for insightful discussion with H. Fujita, and for technical support by K. Ikuo, K. Toyoshima, J. Inoue, T. Haji, K. Hayashi, K. Tanaka, technical assistance division of National Agriculture and Food Research Organization (NARO), NIMS Facilities Maintenance Group, Photron Limited, Photonic Lattice, Inc., Teledyne FLIR LLC, JASCO Corporation, INNOMEDICS Medical Instruments Inc., Covidien Japan Inc., CHEST M.I., INC., and IR System Co., Ltd. This work was supported by Advanced Research Infrastructure for Materials and Nanotechnology in Japan (ARIM) of the Ministry of Education, Culture, Sports, Science and Technology (MEXT). Proposal Number JPMXP25NM5181.

**Disclosures.** The authors declare no conflict of interest.

**Data availability.** Data underlying the results presented in this paper are not publicly available at this time but may be obtained from the authors upon reasonable request.

**Supplemental document.** See Supplement 1 for supporting content.